\documentclass[prx,twocolumn,floatfix,superscriptaddress,longbibliography]{revtex4-1}
\usepackage{tikz}
\usepackage{algpseudocode}
\usepackage[ruled]{algorithm}
\usepackage{mathtools}
\usetikzlibrary{shapes,arrows,positioning}
\usepackage{standalone}
\usepackage{amssymb,amsmath,amstext}
\usepackage{graphicx}
\usepackage{epstopdf}
\usepackage{color}
\usepackage{appendix}
\usepackage[T1]{fontenc}
\usepackage{bbold}
\usepackage{bbm}
\usepackage{float}
\usepackage{latexsym}
\usepackage{xr-hyper}
\usepackage[colorlinks=true,citecolor=blue,linkcolor=magenta]{hyperref}
\usepackage[latin1]{inputenc}

\usepackage{hhline}
 
\usetikzlibrary{shapes,arrows}
\renewcommand{\vec}[1]{\boldsymbol{#1}}  

\long\def\ca#1\cb{} 

\newcommand{\abs}[2][]{#1| #2 #1|}

\newcommand{\avg}[1]{\langle #1\rangle }

\newcommand{\fattheta}{\boldsymbol{\theta}}
\newcommand{\tot}{\text{tot}}

\newcommand{\ave}[1]{\langle #1\rangle}               

\newcommand{\Var}{\text{Var}}

\DeclareMathOperator*{\argmax}{arg\,max}

\renewcommand{\vec}[1]{\boldsymbol{#1}}  

\newcommand{\Eh}{\widehat{E}}

\makeatletter
\newcommand{\justified}{%
  \rightskip=10pt $\,$%
  \leftskip=10pt }
\makeatother
   \makeatletter
     \renewcommand\@make@capt@title[2]{%
      \@ifx@empty\float@link{\@firstofone}{\expandafter\href\expandafter{\float@link}}%
       {\textbf{#1}}\@caption@fignum@sep#2\quad}%
     \makeatother
 
\makeatletter 
\renewcommand{\fnum@figure}{\textbf{Figure~\thefigure}}
\makeatother

\newtheorem{proposition}{Proposition}

\begin{document}

\title{Operator Sampling for Shot-frugal Optimization in Variational Algorithms}

\author{Andrew Arrasmith}
\email{aarrasmith@lanl.gov}
\affiliation{Theoretical Division, MS B213, Los Alamos National Laboratory, Los Alamos, NM 87545, USA.}

\author{Lukasz Cincio} 
\affiliation{Theoretical Division, MS B213, Los Alamos National Laboratory, Los Alamos, NM 87545, USA.}

\author{Rolando D. Somma} 
\affiliation{Theoretical Division, MS B213, Los Alamos National Laboratory, Los Alamos, NM 87545, USA.}

\author{Patrick J. Coles} 
\affiliation{Theoretical Division, MS B213, Los Alamos National Laboratory, Los Alamos, NM 87545, USA.}


\begin{abstract}
Quantum chemistry is a near-term application for quantum computers. This application may be facilitated by variational quantum-classical algorithms (VQCAs), although a concern for VQCAs is the large number of measurements needed for convergence, especially for chemical accuracy. Here we introduce a strategy for reducing the number of measurements (i.e., shots) by randomly sampling operators $h_i$ from the overall Hamiltonian $H = \sum_i c_i h_i$. In particular, we employ weighted sampling, which is important when the $c_i$'s are highly non-uniform, as is typical in chemistry. We integrate this strategy with an adaptive optimizer developed recently by our group to construct an improved optimizer called Rosalin (Random Operator Sampling for Adaptive Learning with Individual Number of shots). Rosalin implements stochastic gradient descent while adapting the shot noise for each partial derivative and randomly assigning the shots amongst the $h_i$ according to a weighted distribution. We implement this and other optimizers to find the ground states of molecules H$_2$, LiH, and BeH$_2$, without and with quantum hardware noise, and Rosalin outperforms other optimizers in most cases. 
\end{abstract}
\maketitle

\section{Introduction}

The variational quantum eigensolver (VQE) is a potential tool for elucidating the electronic structure of molecules and materials~\cite{peruzzo2014VQE}. VQE and other similar variational quantum-classical algorithms (VQCAs)~\cite{farhi2014QAOA, johnson2017qvector, romero2017quantum, larose2018, arrasmith2019variational, cerezo2019variational, jones2019variational, yuan2018theory, li2017efficient, kokail2019self, Khatri2019quantumassisted, jones2018quantum, heya2018variational, endo2018variational,sharma2019noise, carolan2019variational,yoshioka2019variational,bravo-prieto2019,xu2019variational,mcardle2019variational,cirstoiu2019variational,otten2019noise,LubaschVariational20,verdon2019quantum,bravo2019quantum,cerezo2020variational} have been proposed as methods to make use of near-term quantum computers. VQCAs efficiently evaluate a cost function on a quantum computer while optimizing the cost value using a classical computer. Important results have been obtained for the ``quantum portion'' of VQCAs, such as efficient gradient evaluation~\cite{mitarai2018quantum,Schuld2019}, scaling analysis for gradients~\cite{mcclean2018barren,cerezo2020cost}, resilience to certain types of noise~\cite{sharma2019noise,mcclean2016theory}, and reducing measurements by finding commuting operator subsets~\cite{Jena2019,Izmaylov2019,Yen2019,Gokhale2019,Crawford2019,Gokhale2019-2, huggins2019efficient}.



However, to realize the full potential of VQCAs, it is not enough to focus only on the quantum part of these algorithms. One needs a powerful classical optimizer. 
Certain issues arise in VQCAs that are not common in classical algorithms, implying that standard off-the-shelf classical optimizers may not be best suited to VQCAs. For example, multiple runs of quantum circuits are required to reduce the effects of shot noise on cost evaluation. Furthermore, applications like VQE require the measurement of large sets of non-commuting operators, significantly increasing the number of shots needed to obtain a given shot noise~\cite{troyer2015}. An additional complication is that quantum hardware noise flattens the training landscape~\cite{sharma2019noise}. Ideally, for VQCAs, one should design an optimizer that can handle both shot noise and hardware noise.  

Some recent works have focused on classical optimizers~\cite{verdon2019learning, wilson2019optimizing, nakanishi2019, parrish2019, stokes2019quantum, kubler2019adaptive, sweke2019, zhang2019collective,koczor2019quantum,lavrijsen2020classical}. One trend that has emerged is gradient-based optimizers, which are motivated by a result that gradient information improves convergence~\cite{harrow2019}. This approach brings with it the challenge (i.e., the large number of shots required) of potentially needing to estimate many partial derivatives of a function that is a sum over expectation values of many possibly non-commuting observables. 

As a result, our group~\cite{kubler2019adaptive} as well as Sweke et al.~\cite{sweke2019} have recently investigated shot-frugal gradient descent for VQCAs. Specifically, we introduced an optimizer, called iCANS (individual Coupled Adaptive Number of Shots), which outperformed off-the-shelf classical optimizers such as Adam~\cite{Kingma2015} 
for variational quantum compiling and VQE tasks~\cite{kubler2019adaptive}. The key feature of iCANS is that it maximizes the expected gain per shot by frugally adapting the shot noise for each individual partial derivative. 

In this article, we take shot-frugal optimization to the next level. In VQE and other VQCAs, it is common to express the cost function $C = \avg{H}$ as the expectation value of a Hamiltonian $H$ that is expanded as a weighted sum of directly measurable operators $\{ h_i \}_i$:
\begin{equation}
\label{eqn1}
    H=\sum_{i=1}^N c_i h_i.
\end{equation} 
Then $C$ is computed from estimations of each expectation $\avg{h_i}$, which is obtained from many shots. In this work, we propose to randomly assign shots to the $h_i$ operators according to a weighted probability distribution (proportional to $|c_i|$). We prove that this leads to an unbiased estimator of the cost $C$, even when the number of shots is extremely small (e.g., just a single shot). This allows one to unlock a level of shot-frugality for unbiased estimation that simply cannot be accessed without operator sampling. In addition, the randomness associated with operator sampling can provide a means to escape from local minima of $C$. We note that randomly sampling the $h_i$ terms was also examined in Ref.~\cite{sweke2019} although their approach is different from ours (as discussed in Sec.~\ref{sec:single-rand}), and it was also employed by Campbell in the context of dynamical simulation and phase estimation~\cite{campbell2019random}.

A combination of the new sampling strategy with iCANS leads to our main result, which is an improved optimizer for VQCAs that we call Rosalin (Random Operator Sampling for Adaptive Learning with Individual Number of shots). Rosalin retains the crucial feature of maximizing the expected gain (i.e., cost reduction) per shot. We analyze the potential of Rosalin by applying it to VQE for three molecules, namely H$_2$, LiH, and BeH$_2$, and compare
its performance with that of other optimizers. In cases with more than a few terms in the Hamiltonian, Rosalin outperforms all other optimizer and sampling strategy combinations considered. Hence, we believe Rosalin should be regarded as the state-of-the-art method for any application that is concerned about shot frugality.

\section{Results}

\subsection{Variances of Estimation Strategies}\label{sec:analytical}



In what follows, we compare various strategies (in terms of their variances) for estimating expectation values with a finite total number of shots, $s_\tot$. For this purpose, we denote $\Eh$ as the estimator for $\avg{H}$ and $\widehat{\mathcal{E}_i}$ as the estimator for $\avg{h_i}$, where
\begin{equation}
    \label{eq:gen_expectation}
   \Eh = \sum_{i=1}^N c_i \widehat{\mathcal{E}_i} \,,\quad\text{with}\quad \widehat{\mathcal{E}_i}= \frac{1}{\text{E}[s_i]}\sum_{j=1}^{s_i} r_{ij}\,.
\end{equation}
Here, $s_i$ is the number of shots allocated to the measurement of $h_i$. Note that $s_i$ may be a random variable. As we will work in terms of the total shot budget for the estimation, $s_\tot$, we impose $\sum_{i=1}^Ns_i=s_\tot$. Also, each $r_{ij}$ is an independent random variable associated with the $j$-th measurement of $h_i$. $\text{E}[\cdot]$ denotes the expectation value and we will assume $\text{E}[s_i]>0$ for all $i$. We now state two useful propositions about this estimator.
\begin{proposition}
\label{prop1}
$\Eh$ is an unbiased estimator for $\langle H \rangle$.
\end{proposition}
See Sec.~\ref{App:bias} for a proof of Prop.~\ref{prop1}.

We remark that if $\text{E} [s_{i}]=0$ for any operator, $\Eh$ becomes undefined. However, one could choose to resolve this by modifying \eqref{eq:gen_expectation} to exclude such operators (which we index with the set $\mathcal{I}$) in the operator sum, as this is essentially a statement that one will choose not to measure those operators. Doing this gives
\begin{align}
    \text{E}[\widehat{E}] =\avg{H}-\sum_{i'\in \mathcal{I}} c_{i'} \avg{h_{i}}.
\end{align}
We therefore have that  $\text{E}[\widehat{E}]\ne\left\langle H \right\rangle$ unless $\sum_{i'\in \mathcal{I}} c_{i'} \avg{h_{i'}}=0$. Hence, in the absence of special symmetries or vanishing coefficients, the estimator becomes biased. This justifies our assumption that $\text{E}[s_i]>0$ for all $i$ to achieve an unbiased estimator.

\begin{proposition}
\label{prop2}
The variance of $\widehat{E}$ is
\begin{align}\label{eq:gen_variance}
    {\rm Var}(\widehat{E})= &\sum_{i=1}^N \frac {c_i^2}{{\rm E}[s_{i}]} \sigma_i^2\nonumber\\
    &+\sum_{i, i'}\frac{c_ic_{i'}\langle h_i \rangle \langle h_{i'} \rangle }{{\rm E}[s_i]{\rm E}[s_{i'}]}{\rm Cov}[s_i,s_{i'}]\;,
\end{align}
where $\sigma_i^2=\langle h_i^2 \rangle-\langle h_i \rangle^2$ is the quantum mechanical variance of $h_i$ in the given state and ${\rm Cov}[s_i,s_{i'}]={\rm E}[s_is_{i'}]-{\rm E}[s_i] {\rm E}[s_{i'}]$ are the entries of the covariance matrix associated with the $s_i$'s.. 
\end{proposition}
See Sec.~\ref{App:Var} for a proof of Prop.~\ref{prop2}.


We note that in this formalism each $h_i$ operator can either be a unitary operator (such as tensor products of Pauli operators) or, more generally, a weighted sum over a commuting set of unitary operators. For simplicity, we will work with normalized versions of these operators ($h_i'=h_i/\|h_i\|$) and absorb the norm into the coefficients ($c_i'=c_i\|h_i\|$) so that $c_i h_i=c_i' h_i'$. For the remainder of this article, we will assume that all $h_i$'s are normalized in this way and drop the primes. Additionally, we define $M=\sum_{i=1}^N|c_i|$ for convenience.


\subsubsection{Uniform Deterministic Sampling}\label{sec:uniform}

The simplest approach to estimating $\langle H\rangle$ with a finite total number of shots $s_{\tot}$ is to simply divide the number of shots equally among the $N$ terms. That then leads us to $s_i=s_\tot/N$. Note that since we need to work with positive integer numbers of shots, this strategy is only valid for $s_\tot=nN$ for some positive integer $n$. When working with optimization environments where other values of $s_\tot$ may be requested by the method, we resolve the disparity by instead using $s_\tot'=n'N$ shots, where $n'=\left\lfloor s_\tot/N\right\rfloor$. We will refer to this strategy below as Uniform Deterministic Sampling (UDS).

With this deterministic strategy, ${\rm E}[s_i]=s_\tot/N$ for all $i$, and ${\rm Cov}[s_i,s_{i'}]=0$ for all $i,i'$. From~\eqref{eq:gen_variance} we then have
\begin{equation}\label{eq:uni-var}
    \Var\left(\widehat{E}\right)=\frac{N}{s_{\text{tot}}}\sum_{i=1}^N\abs{ c_i}^2 \sigma_i^2.
\end{equation}

We note that this strategy represents the optimal allocation of shots in the special case where $\sigma_i\propto 1/\abs{c_i}$~\cite{rubin2018application}.


\subsubsection{Weighted Deterministic Sampling}\label{sec:weighted}

To improve the shot frugality of reaching a given precision, it has been proposed~\cite{troyer2015,rubin2018application} that the number of shots $s_i$ allocated to the measurement of each operator $h_i$ should be proportional to the magnitude of the coefficients $c_i$. That is, the shots would be deterministically proportioned so that:
\begin{equation}
    s_i= s_\tot\frac{\abs{c_i}}{M}.
\end{equation}
Note that for physical Hamiltonians $s_\tot \abs{c_i}/{M}$ will often not be an integer. When this occurs we again take the floor:
\begin{equation}
    \label{eq:weighted_alloc}
    s_i=\left\lfloor s_\tot\frac{\abs{c_i}}{M}\right\rfloor,
\end{equation}
which also redefines the total number of shots used as $s_\tot'=\sum_{i=1}^N\lfloor s_\tot{\abs{c_i}}/M\rfloor$. We will refer to this strategy below as Weighted Deterministic Sampling (WDS). 

From~\eqref{eq:gen_variance}, the variance of the estimator for this deterministic strategy (neglecting any corrections due to $s_{\rm tot}|c_i|/M$ not being integer) is
\begin{equation}\label{eq:weighted-var}
    \Var\left(\widehat{E}\right)=\frac{{M}}{s_{\text{tot}}}\sum_{i=1}^N |c_i| \sigma_i^2\,.
\end{equation}

For equal magnitude coefficients, this method reduces to the UDS approach above, while when the $\sigma_i$'s are equal in magnitude this strategy becomes optimal~\cite{rubin2018application}. In the case of performing VQE for chemical Hamiltonians, we empirically find that for both random states as well as low energy states there is a greater variation in the $\abs{c_i}$'s than in the $\sigma_i$'s, and so WDS tends to perform better than UDS.

\subsubsection{Weighted Random Sampling}\label{sec:weighted-rand}

The above deterministic frameworks have a hard floor on the number of shots needed for any unbiased estimate, and this floor increases with $N$ as those methods must measure all operators at least once. This is a crucial point: deterministic methods cannot be unbiased when the number of shots is below some threshold, and hence this limits the shot frugality of these methods. In general, this shot floor is derived from demanding that $\min_i\,s_i>0$. For the specific case of WDS this floor is
\begin{equation}
    \label{eq:ShotFloor}
 s_{\text{floor}} =\left \lceil \frac{M}{\min_i \abs{c_i}} \right \rceil  \le  s_{\text{tot}} \,.
\end{equation}

To lower this floor, one can introduce randomness. With randomness, an unbiased estimator for $\avg{H}$ can be computed with as little as a single shot. Randomness, therefore, unlocks a new level a shot frugality. While this shot frugality naturally leads to noisy gradient estimates, it has been demonstrated for VQCAs that stochastic gradient descent can be effective even with highly noisy gradient estimates~\cite{kubler2019adaptive,sweke2019}.



We therefore propose a weighted random sampling (WRS) method. This sampling can be accomplished by choosing which $h_i$ terms to measure by drawing from a multinomial probability distribution where the probability $p_i$ of measuring $h_i$ is proportional to the magnitude of the coefficient $c_i$:
\begin{equation}
    p_i=\frac{\abs{c_i}}{M}.
\end{equation}
Note that here $\text{E} [s_i]=p_is_{\tot}$. The variance of this estimator then follows from plugging the variance and covariance of the multinomial distribution into~\eqref{eq:gen_variance}:
\begin{align}\label{eq:rand-var}
    \Var\left(\widehat{E}\right)=&\frac{{M}}{s_{\text{tot}}}\sum_{i=1}^N |c_i| \avg{h_i^2}-\frac{\avg{H}^2}{s_{\text{tot}}}\nonumber \\
    =&\frac{{M}}{s_{\text{tot}}}\sum_{i=1}^N |c_i| \sigma_i^2\nonumber \\
    &+\frac{{M}}{s_{\text{tot}}}\sum_{i=1}^N |c_i| \avg{h_i}^2-\frac{\avg{H}^2}{s_{\text{tot}}}\,.
\end{align}
When allowed to take many shots this procedure is very similar to the weighted, deterministic method above while also extending to the few shots, high variance regime. The price for this extension is the additional two terms added to the variance, the sum of which is always non-negative as it represents an expectation value for a (positive-semidefinite) covariance matrix. We note that these additional terms do not alter the ($1/s_\tot$) scaling of the deterministic case.

\subsubsection{Weighted Hybrid Sampling}\label{sec:weighted-hyb}
There is a middle ground between the deterministic and stochastic weighted sampling procedures listed above: we can allocate some of the shots according to each method. To this end, we divvy up the shots as follows. If $s_{\tot} \ge s_{\text{floor}}$, we first assign shots according to the WDS strategy (i.e.,~\eqref{eq:weighted_alloc}) and then assign any leftover shots randomly:
\begin{equation}
    s_{\text{rand}}=s_{\tot}-\sum_{i=1}^N\lfloor s_\tot{\abs{c_i}}/M\rfloor\, ,
\end{equation}
where $s_{\text{rand}}$ is the number of shots allocated randomly according to the WRS strategy. If instead $s_{\tot} < s_{\text{floor}}$ we set $s_{\text{rand}}=s_{\tot}$ and allocate all of the shots randomly. This gives the following variance:
\begin{align}\label{eq:hyb-var}
    \Var\left(\widehat{E}\right)=&\frac{{M}}{s_{\text{tot}}}\sum_{i=1}^N |c_i| \sigma_i^2\nonumber \\
    &+\frac{s_{\text{rand}}M}{s_{\text{tot}}^2}\sum_{i=1}^N |c_i| \avg{h_i}^2-\frac{s_{\text{rand}}\avg{H}^2}{s_{\text{tot}}^2}\,.
\end{align}
This formula follows from~\eqref{eq:weighted-var} and~\eqref{eq:rand-var} by standard properties of the variance of the sum of two independent random variables. We note that this variance is no smaller than~\eqref{eq:weighted-var}. We will refer to this strategy as Weighted Hybrid Sampling (WHS).

Since $s_{\text{rand}}$ is bounded above, the two terms that are added to the WDS formula scale as 1/$s_{\text{tot}}^2$, and thus will not contribute significantly to the variance at high numbers of shots. This method is therefore well suited to both the low and high number of shots regimes, making it very useful in the context of optimization methods.

\begin{figure*}[!ht]
    \centering
    \includegraphics[width=2\columnwidth]{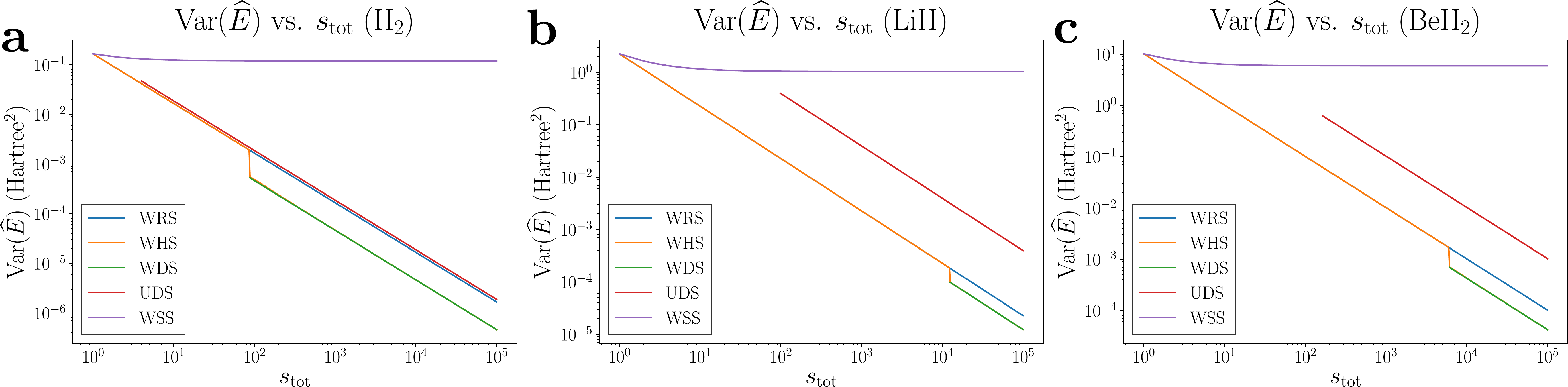}
    \caption{Variances of the estimator $\Eh$ with the different sampling strategies for the low energy states of different Hamiltonians. Panel \textbf{a} shows the results for a low energy state for H$_2$ generated by optimizing the angles in the ansatz in Fig.~\ref{fig:ansatz} with $D=1$. Panel \textbf{b} shows the same but for LiH and with $D=2$. Panel \textbf{c} again shows the same but for BeH$_2$ and with $D=2$.}
    \label{fig:Var}
\end{figure*}

\subsubsection{Randomly Selecting Only One Term}\label{sec:single-rand}

A different method of randomly sampling was recently proposed where instead of distributing the shots at random, one randomly selects a single $h_i$ and uses all of the shots to measure that $h_i$~\cite{sweke2019}. As shown therein, this method provides an unbiased estimator for $\avg{H}$. In our notation, the estimator for this method is given by
\begin{equation}
    \label{eq:single_expectation}
   \widehat{E} =  \frac{c_i}{s_\text{tot}}\sum_{j=1}^{s_\text{tot}} r_{ij}\,,
\end{equation}
for some $i\in \{1,2,\dots,N\}$ chosen with probability $p_i$. Though the authors of~\cite{sweke2019} focused on the uniform case where $p_i=1/N$, they also comment that one could use a weighted approach like the one used in the WRS and set $p_i=\abs{c_i}/M$. 

If one uses this weighted approach, which we will refer to as Weighted Single Sampling (WSS), the variance of their estimator is
\begin{align}\label{eq:single-var}
    \Var\left(\widehat{E}\right)=&\frac{M}{s_{\text{tot}}}\sum_{i=1}^N |c_i| \sigma_i^2\nonumber \\
    &+M\sum_{i=1}^N |c_i| \avg{h_i}^2-\avg{H}^2\,.
\end{align}
As with the variance of WRS, this follows from plugging in the variance of the multinomial distribution into~\eqref{eq:gen_variance} but differs as we only take a single draw.

From comparing~\eqref{eq:rand-var} and~\eqref{eq:single-var}, one can see that measuring only one operator adds a floor to the variance as the additional terms no longer scale as ($1/s_\tot$). As mentioned in~\cite{sweke2019}, this method can be extended to distributing the shots among a subset of the $h_i$'s which improves the situation but cannot recover the $1/s_\tot$ scaling unless we include all $h_i$'s.

\begin{figure}[]
    \centering
    \includegraphics[width=\columnwidth]{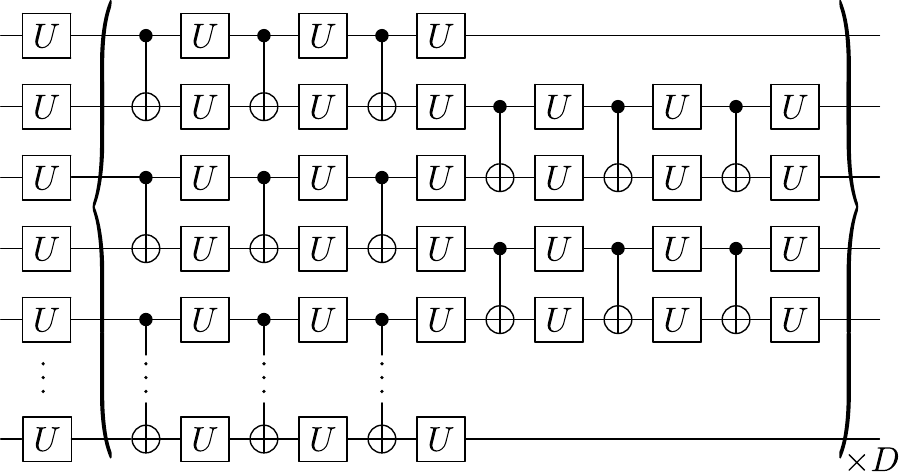}
    \caption{Structure of the quantum circuit ansatz used in our numerics. The block of gates inside the curly braces is repeated $D$ times to provide different depth ansatzes. Each $U$ gate represents a general single qubit unitary and is the composition of a z-rotation, a y-rotation, and a z-rotation. Each angle in these rotations is varied independently.}
    \label{fig:ansatz}
\end{figure}

\subsubsection{Numerical Comparison of Variances}\label{sec:numerical}

As the variance from any sampling strategy will depend both on the state and the Hamiltonian, we now consider the variance for states of interest for quantum chemistry. Specifically, we now compare them numerically for the low energy states found at the end of a VQE procedure in Fig.~\ref{fig:Var}. We employ the ansatz structure described in Fig.~\ref{fig:ansatz} with the Hamiltonians for H$_2$, LiH, and BeH$_2$ used in \cite{kandala2017}.


As shown in Fig.~\ref{fig:Var}, for all cases, WDS is the best at high values of $s_\tot$ but does not come into play until we are allocating many shots, especially for LiH and BeH$_2$. WRS and WHS are identical for small $s_\tot$, and perform the best there. Once WDS becomes relevant, the variance of WHS jumps to meet it and then stays close, showing an advantage over WRS. For all cases, the WRS, WHS, and WDS (when relevant) give smaller variances than UDS, though we note that UDS becomes possible at fewer shots than WDS. Due to the variance floor of WSS, it is typically not competitive with the other strategies.

\subsection{The Rosalin Method}\label{sec:Rosalin}

In order to formulate an optimizer geared towards chemistry applications, we combine the shot-frugal optimizer iCANS~\cite{kubler2019adaptive} (in particular the more aggressive variant referred to as iCANS1 in that paper) with the WRS and WHS strategies described above. We refer to the resulting method as Rosalin (Random Operator Sampling for Adaptive Learning with Individual Number of shots). We present the random and hybrid operator sampling methods in Algorithm \ref{alg:expectation}, and review the iCANS method in Algorithm \ref{alg:iCANS} in the Sec.~\ref{sec:iCANS}. Together, these methods compose the Rosalin approach to VQE and other VQCA problems. We refer to the WRS version of Rosalin as Rosalin1 and the WHS version as Rosalin2.

\begin{figure}[!t]
\begin{algorithm}[H]
\begin{algorithmic}[1]
\let\oldReturn\Return
\renewcommand{\Return}{\State\oldReturn}
\Procedure{$Estimate\_H$}{$\fattheta,  s_\tot,\{c_i\},\{h_i\}$}
    \State initialize: $\vec{\Eh} \gets (0,...,0)^T$, $\ell \gets 0$, $\vec{s} \gets (0 ,... ,0)^T$
    \For{$i \in [1,...,N]$}
        \State $p_i \gets\frac{|c_i|}{\sum_{i'}|c_{i'}|}$
    \EndFor
    \If{$Hybrid$ and $\lfloor \min_i( p_is_\tot)\rfloor >0$}
        \For{$i \in [1,...,N]$}
            \State $s_i \gets \lfloor p_i s_\tot\rfloor$
        \EndFor
        \State $s_{\text{det}} \gets \sum_i \vec{s}_i$
    \Else
        \State $s_{\text{det}} \gets 0$
    \EndIf
    \State $s_{\text{rand}} \gets s_\tot-s_{\text{det}}$
    \State $\vec{m} \sim \text{Multinomial}(s_{\text{rand}},\vec{p})$
    \For{$j \in [1,...,s_{\text{rand}}]$}
        \State $s_{m_j} \gets s_{m_j} + 1$
    \EndFor
    \For{$i \in [1,...,N]$}
        \For{$j \in [1,...,s_i]$}
            \State $r \gets Measure(\fattheta,h_i)$
            \State $\ell \gets \ell +1$
            \State $\Eh_\ell \gets c_i r/p_i$
        \EndFor

    \EndFor
    
    \Return $\vec{\Eh}$
\EndProcedure
\end{algorithmic}
\caption{\justified{The function $Estimate\_H$ which, given a parameter vector $\vec{\theta}$, a shot budged $s_\tot$, and the sets of coefficients and operators, $\{c_i\}$ and $\{h_i\}$, returns a vector of single-shot estimates ($\vec{\Eh}$) of $\langle H \rangle$ for Rosalin using either the WHS or the WRS strategy, depending on the Boolean $Hybrid$ flag. The function $Measure(\fattheta,h_i)$ represents a measurement of the operator $h_i$ on a state prepared by the circuit parametrized by $\fattheta$.}}
\label{alg:expectation}
\end{algorithm}
\end{figure}

\begin{figure*}[!ht]
    \centering
    \includegraphics[width=2\columnwidth]{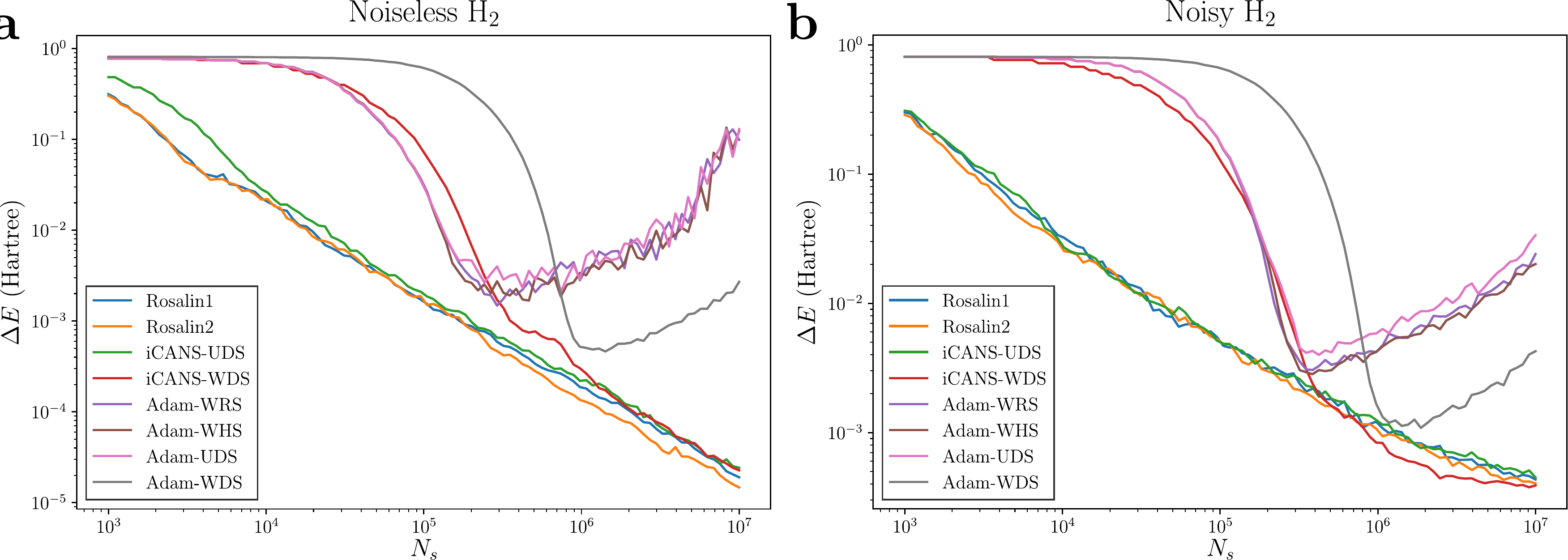}
    \caption{Average energy above the ground state ($\Delta E$) for the H$_2$ molecular Hamiltonian as a function of the total number of shots ($N_s$) expended during the optimization procedure. The energy is calculated exactly using the parameters found with the stochastic optimization. Each of the 4 Pauli product operators in the Hamiltonian description were measured separately. Panels \textbf{a} and \textbf{b} show the results for optimizing without and with machine noise (respectively). Both cases were optimized with the ansatz in Fig.~\ref{fig:ansatz} with $D=1$.}
    \label{fig:H2}
\end{figure*}
\begin{figure*}[!ht]
    \centering
    \includegraphics[width=2\columnwidth]{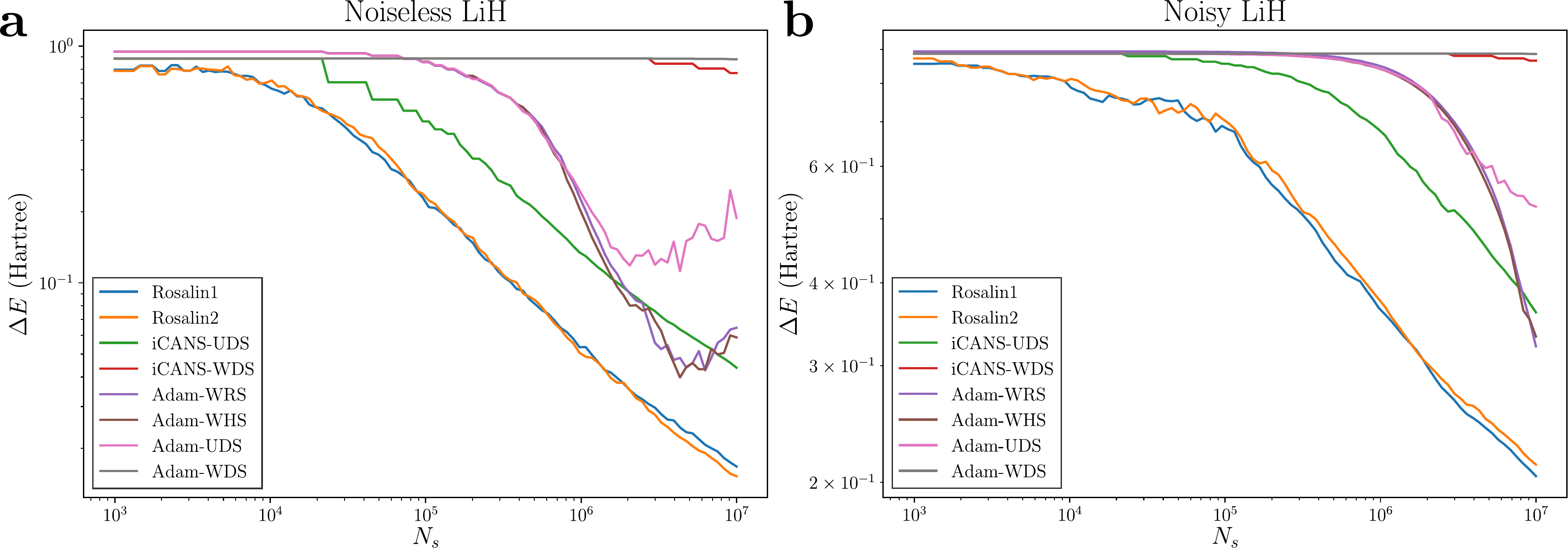}
    \caption{Average energy above the ground state ($\Delta E$) for the LiH molecular Hamiltonian as a function of the total number of shots ($N_s$) expended during the optimization procedure. The energy is calculated exactly using the parameters found with the stochastic optimization. Each of the 99 Pauli product operators in the Hamiltonian description were measured separately. Panels \textbf{a} and \textbf{b} show the results for optimizing without and with machine noise (respectively). Both cases were optimized with the ansatz in Fig.~\ref{fig:ansatz} with $D=2$.}
    \label{fig:LiH}
\end{figure*}
\begin{figure*}[!t]
    \centering
    \includegraphics[width=2\columnwidth]{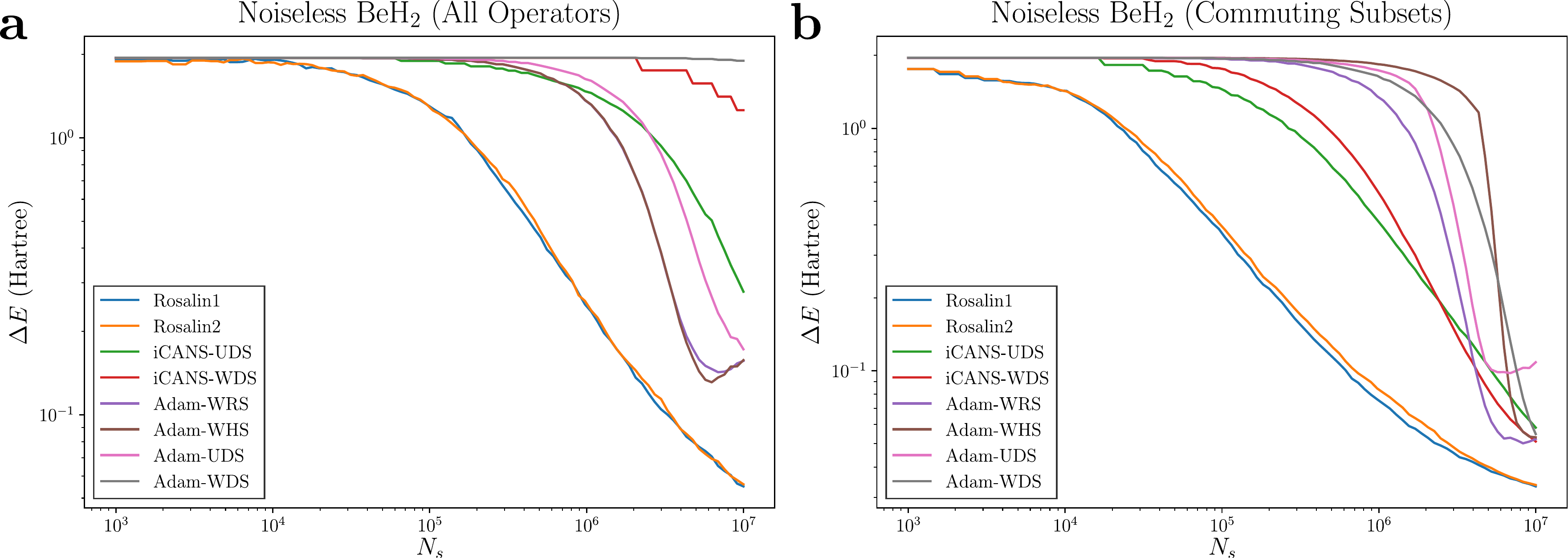}
    \caption{Average energy above the ground state ($\Delta E$) for the BeH$_2$ molecular Hamiltonian as a function of the total number of shots ($N_s$) expended during the optimization procedure. The energy is calculated exactly using the parameters found with the stochastic optimization. Panel \textbf{a} shows the results while measuring each of the 164 Pauli product operator separately while panel \textbf{b}  shows the results when simultaneously measuring operators within the 44 commuting subsets of Pauli product operators. Both cases were optimized with the ansatz in Fig.~\ref{fig:ansatz} with $D=2$.}
    \label{fig:BeH2}
\end{figure*}
\subsection{Numerical Optimization Comparison} \label{sec:Implementations}
We now compare the performance of different sampling strategies with two optimizers on the optimization of the molecular Hamiltonians used in \cite{kandala2017}: H$_2$, LiH, and BeH$_2$. For each Hamiltonian, we consider eight different approaches arising from using either iCANS or Adam with four different sampling strategies: WRS, WHS, WDS, and UDS.

In Fig.~\ref{fig:H2} and Fig.~\ref{fig:LiH} we show the average performance of these optimization strategies for the H$_2$ and LiH molecular VQE problems versus the total number of shots expended. We perform this comparison both with and without simulated hardware error (both versions are limited by the finite statistics). The noise model used here was based on the noise profile of IBM's Melbourne processor~\cite{IBMQ14} as retrieved by Qiskit function calls~\cite{gadi_aleksandrowicz_2019_2562111}. In Fig.~\ref{fig:BeH2} we instead compare the performance of these optimization strategies for minimizing the energy of BeH$_2$ both with considering each operator separately (as in the H$_2$ and LiH cases above) and using the commuting subsets chosen in~\cite{kandala2017}. All instances using the Adam optimizer were given the larger of one hundred shots or the method's shot floor for the given Hamiltonian for each expectation value estimated.


\section{Discussion}
\label{sec:Discussion}
In order to achieve practical applications of VQCAs, we will need optimization strategies that can scale well as we consider larger problem sizes. In particular, applying VQE to chemical applications will have to contend with Hamiltonians that are the sum over many directly measurable operators. While simultaneously measuring commuting subsets of these operators can help reduce this difficulty, it is only one part of the answer.

With this challenge in mind, we have introduced the Rosalin method which combines random operator sampling with the iCANS optimizer to achieve greater shot frugality. iCANS is a recently introduced optimizer that attempts to be shot-frugal by dynamically adapting the number of shots (and thus precision) used to estimate each gradient component as part of a stochastic gradient descent~\cite{kubler2019adaptive}. By combining iCANS and the random or hybrid sampling strategies (WRS or WHS, respectively), Rosalin has the ability to make gradient update steps that are very inexpensive (i.e., use few shots) early in the training process when less precision is needed, even for large and complicated Hamiltonians. Additionally, since our random and hybrid sampling schemes retain the standard  $1/s_\tot$ scaling, Rosalin is able to directly increase the precision it uses as needed during the optimization procedure. 

The analytical results of Sec.~\ref{sec:analytical} show how the usual $1/s_{\text{tot}}$ scaling of the variance can be achieved by random sampling strategies, while also allowing for unbiased estimators with far fewer resources. Additionally, they show that such a random sampling strategy should allocate samples across all of the operators $h_i$ (rather than singling out an individual operator) in order to avoid introducing a precision floor. These results about our random and hybrid sampling strategies are central to the shot frugality advantage that we find with Rosalin.

The simulated optimization procedures of Sec.~\ref{sec:Implementations} show that as the size and complexity of the Hamiltonians being considered increases, random sampling procedures such as Rosalin offer greater improvements in the efficiency of performing an optimization. Specifically, for the case of H$_2$ where we had a Hamiltonian with only $4$ terms Fig.~\ref{fig:H2} does not show much of a difference between the sampling strategies. However, once we move on to LiH with $99$ terms we see a marked difference for both the noisy and noiseless optimization in Fig.~\ref{fig:LiH}. This difference is even more pronounced for the larger molecule BeH$_2$ in Fig.~\ref{fig:BeH2}.  Additionally, for this molecule, we find that Rosalin achieves an advantage over the other methods considered even when we work with the $44$ commuting subsets (Fig.~\ref{fig:BeH2}\textbf{b}) rather than measuring the full $164$ terms individually (Fig.~\ref{fig:BeH2}\textbf{a}). 

We note that while our results have concerned unbiased estimators, there may sometimes be a case for using a biased estimator. We focus on unbiased estimators as stochastic optimization methods using a biased estimator would generically be expected to converge to a distribution of parameter values that is not centered about the true minimum. However, if our goal is to compute the energy of a ground state to a fixed accuracy rather than trying to find the true optimal state parameters, we might choose to exclude a set of terms in the Hamiltonian chosen to keep the total energy bias small enough that it is negligible compared to the desired accuracy. The random and hybrid methods we propose here would apply as naturally to such a biasing truncation of a Hamiltonian as to the full Hamiltonian.

Finally, we remark that while Rosalin is an optimization strategy intended for VQCAs, the random and hybrid sampling strategies we propose would also provide a way to potentially achieve less computationally expensive estimates of expectation values for non-variational methods as well. Hence, our work has relevance to traditional quantum algorithms, even in the fault-tolerant quantum computing era.


\section{Methods}

\subsection{Proof of Prop.~\ref{prop1}}\label{App:bias}
 In~\eqref{eq:gen_expectation} we introduced the following expression to estimate $\ave{H}$ with $s_\text{tot}$ shots:
\begin{equation}
\label{eq:gen_exp_form}
   \Eh = \sum_{i=1}^N c_i  \frac{1}{\text{E}[s_i]}\sum_{j=1}^{s_i} r_{ij},
\end{equation}
where $r_{ij}$ is the measurement outcome of the $j$-th measurement of $h_i$ and $s_i$ is the (possibly random) number of shots allocated to the measurement of $h_i$, and we assume $\text{E} [s_i]>0$ for all $i$. We now show that $\text{E}[\widehat{E}]=\left\langle H \right\rangle$  (Prop.~\ref{prop1}). We have:
\begin{align}
    \text{E}[\widehat{E}] =&\text{E}\Bigg[ \sum_{i=1}^N c_i  \frac{1}{\text{E}[s_i]}\sum_{j=1}^{s_i} r_{ij}\Bigg]\nonumber\\
    =& \sum_{i=1}^N c_i  \frac{1}{\text{E}[s_i]}\text{E}\Bigg[\sum_{j=1}^{s_i} r_{ij}\Bigg]\nonumber\\
    =& \sum_{i=1}^N c_i \frac{\text{E}[s_i]}{\text{E}[s_i]}\avg{h_i}\nonumber\\
    =&\avg{H}.
\end{align}
Thus we have shown that if $\text{E} [s_i]>0$ for all $i$, then $\text{E}[\widehat{E}]=\left\langle H \right\rangle$.

\subsection{Proof of Prop.~\ref{prop2}}\label{App:Var}
We now derive the general variance formula~\eqref{eq:gen_variance} to prove Prop.~\ref{prop2}. We have
\begin{align}
      {\rm Var}(\Eh) =&  \sum_{i, i'=1}^N\frac{c_ic_{i'}}{{\rm E}[s_i]{\rm E}[s_{i'}]}{\rm E} [\sum_{j}^{s_i}\sum_{j'}^{s_{i'}} r_{ij}r_{i'j'}] - \langle H \rangle^2 \nonumber\\
      =&\sum_{i, i'=1,i \ne i'}^N\frac{c_ic_{i'}\langle h_i \rangle \langle h_{i'} \rangle }{{\rm E}[s_i]{\rm E}[s_{i'}]}{\rm E}[s_i s_{i'}]\nonumber\\ 
      &+ \sum_{i=1}^N\frac{(c_i)^2 \langle h_i \rangle^2  }{{\rm E}^2[s_i]}{\rm E}[s_i (s_{i}-1)] \nonumber\\
     & + \sum_{i=1}^N\frac{(c_i)^2 \langle h_i^2 \rangle  }{{\rm E}^2[s_i]}{\rm E}[s_i] - \langle H \rangle^2 \nonumber\\
     =& \sum_{i=1}^N \frac {c_i^2}{{\rm E}[s_{i}]} \sigma_i^2  +\sum_{i, i'=1}^N\frac{c_ic_{i'}\langle h_i \rangle \langle h_{i'} \rangle }{{\rm E}[s_i]{\rm E}[s_{i'}]}{\rm E}[s_i s_{i'}]\nonumber\\
     &- \langle H \rangle^2 \nonumber\\
     =&\sum_{i=1}^N \frac {c_i^2}{{\rm E}[s_{i}]} \sigma_i^2 +\sum_{i, i'=1}^N\frac{c_ic_{i'}\langle h_i \rangle \langle h_{i'} \rangle }{{\rm E}[s_i]{\rm E}[s_{i'}]}{\rm Cov}[s_i,s_{i'}] \;,
\end{align}
which is~\eqref{eq:gen_variance}.

\subsection{Remark on Using Prior Variance Information}\label{sec:prior_var}

If one approximately knows the $\sigma_i$'s ahead of time for a quantum state of interest, the proportioning of shots in both the deterministic and stochastic sampling methods can be optimized further to decrease the variance of the estimate~\cite{rubin2018application}. If one has such information available, optimal deterministic distribution of shots is
\begin{equation}
    \label{eq:weighted_alloc_prior}
    s_i=\frac{s_{\tot}\abs{c_i}\sigma_i}{\sum^N_{i'=1}|c_{i'}|\sigma_{i'}}.
\end{equation}
(As in the other deterministic cases discussed, this may not result in an integer $s_i$, in which case once could take the floor of this expression.)
Note that for this estimator to remain unbiased we must demand that at least one shot is allocated to each operator, meaning that this prescription requires some form of regularization when $\sigma_i\to 0$. However, including such prior information makes the variance of the estimator 
\begin{equation}\label{eq:weighted-var-with-info}
    \Var\left(\widehat{E}\right)=\frac{(\sum^N_{i'=1}|c_{i'}|\sigma_{i'})^2}{s_{\text{tot}}},
\end{equation}
which is optimal~\cite{rubin2018application}. 

In the random approach, if we know the $\sigma_i$'s we can instead set the probabilities to be 
\begin{equation}
    p_i=\frac{\abs{c_i}\sigma_i}{\sum_{i'}\abs{c_{i'}}\sigma_{i'}}\,,
\end{equation}
 which gives a variance of
\begin{equation}\label{eq:rand-var-with-info}
    \Var\left(\widehat{E}\right)=\frac{\sum^N_{i'=1}|c_{i'}|\sigma_{i'}}{s_{\text{tot}}}\sum_{i=1}^N |c_i|\frac{ \avg{h_i^2}}{\sigma_i}-\frac{\avg{H}^2}{s_{\text{tot}}}\,.
\end{equation}
Note that this variance diverges when at least one $\sigma_i$ becomes small, meaning that this method is unstable without regularizing the expression with an effective lower bound on the $\sigma_i$'s.

Additionally, we empirically find that during a minimization procedure using the $\sigma_i$'s from the previous iteration to guide the shot allocation for a new iteration performs poorly. Therefore, though such information may be helpful when accurately determining an expectation value (perhaps after an optimization procedure), we do not incorporate it into Rosalin.

\subsection{The Rosalin Optimizer}\label{sec:iCANS}

Algorithm \ref{alg:iCANS} is described below and shows the version of iCANS~\cite{kubler2019adaptive} adapted to Rosalin. We now make a few remarks about the hyperparameters of Rosalin. These include the Lipshitz constant $L$, the maximum learning rate $\alpha$, the running average constant $\mu$, the minimum number of shots per energy estimation $s_{\text{min}}$, and the bias $b$. 

$L$ is the bound on the largest derivative that the energy landscape has and is thus set by the problem Hamiltonian. We recommend setting $L =M$, as suggested in \cite{kubler2019adaptive}. $L$ also gives a bound on the learning rates that can be used, as the iCANS formalism requires that $0<\alpha<2/L$. 

The running average constant $\mu$ is bounded between zero and one. Unlike in other methods with running averages, $\mu$ is only used to control how quickly the number of shots recommended for each gradient component changes. With this in mind, $\mu$ can be set close to one in order to get a smooth increase in the number of shots without directly influencing the parameter update step. We also note that $s_{\text{min}}$ cannot be lower than $2$ for the variance $\vec{S}_\ell$ to be well defined (see line 22 of Algorithm \ref{alg:iCANS}). Finally, the bias $b$ is introduced to act as a regularizer and thus should be positive but small compared to the expected size of the variances.
\begin{figure}[ht]
\begin{algorithm}[H]
\begin{algorithmic}[1]
\let\oldReturn\Return
\renewcommand{\Return}{\State\oldReturn}
\Statex \textbf{Input:} Learning rate $\alpha$, starting point $\fattheta_0$, min number of shots per estimation $s_{\min}$, number of shots that can be used in total $M$, Lipschitz constant $L$, running average constant $\mu$, bias for gradient norm $b$, and Hamiltonian expressed as a list of coefficients $c_i$ and operators $h_i$.
\State initialize: $\fattheta \gets \fattheta_0 $, $s_{\text{used}} \gets 0$,
$\vec{s} \gets (s_{\min} ,... ,s_{\min})^T$, $\vec{\chi} \gets (0,...,0)^T$,
$\vec{\xi} \gets (0,...,0)^T$, $k\gets 0$
\While{$s_{\text{used}} < M$}
\State $s_{\text{used}} \gets s_{\text{used}} + 2 \sum_\ell s_\ell$
\For{$ \ell \in [1,...,d]$}
    \State $g_\ell, S_\ell \gets iEvaluate(\fattheta, s_\ell,\ell,\{c_i\},\{h_i\})$
    \State $\xi_\ell' \gets \mu \xi_\ell' + (1-\mu) S_\ell$
    \State $\chi_\ell' \gets \mu \chi_\ell' + (1-\mu) g_\ell$ 
    \State $\xi_\ell \gets \xi_\ell'/(1-\mu^{k+1})$
    \State $\chi_\ell \gets  \chi_\ell'/(1-\mu^{k+1})$ 

            \State $\theta_\ell \gets \theta_\ell - \alpha g_\ell$
    \State $s_\ell \gets \left\lceil\frac{2L\alpha}{2-L\alpha} \frac{\xi_\ell}{\chi_\ell^2 + b \mu^k}\right\rceil$
    \State $\gamma_\ell \gets \frac{1}{s_\ell} \left[\left(\alpha - \frac{L\alpha^2}{2}\right) \chi_\ell^2 - \frac{L \alpha^2}{2 s_\ell} \xi_\ell\right]$
\EndFor
\State $s_{\max} \gets s_{\argmax_\ell \gamma_\ell}$ 
\State $\vec{s} \gets clip(\vec{s}, s_\text{min} , s_\text{max} )$ 
\State $k\gets k + 1$
\EndWhile
\Procedure{$iEvaluate$}{$\fattheta,  s_\tot,\ell,\{c_i\},\{h_i\}$}
  \State $\vec{\Eh}^+\gets Estimate\_H(\fattheta+ \frac{\pi}{2}\hat{e}_\ell,s_\tot,\{c_i\},\{h_i\})$ 
  \State $\vec{\Eh}^-\gets Estimate\_H(\fattheta- \frac{\pi}{2}\hat{e}_\ell,s_\tot,\{c_i\},\{h_i\})$
  \State $g_\ell \gets \sum_{j=1}^{s_\tot} (\Eh^+_j-\Eh^-_j)/(2s_\tot)$
  \State $S_\ell \gets \sum_{j=1}^{s_\tot} [((\Eh^+_j-\Eh^-_j)/(2))^2-\vec{g_\ell}^2]/(s_\tot-1)$
  \Return $g_\ell,S_\ell$
\EndProcedure
\end{algorithmic}
\caption{\justified{The optimization loop for Rosalin. The function $Estimate\_H$ which returns a vector containing single-shot estimates of $\langle H \rangle$ for Rosalin is described in Algorithm~\ref{alg:expectation}. }}
\label{alg:iCANS}
\end{algorithm}
\end{figure}

\begin{acknowledgements}

AA, LC, RDS, and PJC acknowledge support from the LDRD program at LANL. PJC also acknowledges support from the LANL ASC Beyond Moore's Law project. LC, RDS, and PJC were also supported by the U.S. Department of Energy, Office of Science, Office of Advanced Scientific Computing Research, under the Accelerated Research in Quantum Computing (ARQC) program. Los Alamos National Laboratory is managed by Triad National Security, LLC, for the National Nuclear Security Administration of the U.S. Department of Energy under Contract No. 89233218CNA000001.

\end{acknowledgements}


\begin{thebibliography}{57}%
\makeatletter
\providecommand \@ifxundefined [1]{%
 \@ifx{#1\undefined}
}%
\providecommand \@ifnum [1]{%
 \ifnum #1\expandafter \@firstoftwo
 \else \expandafter \@secondoftwo
 \fi
}%
\providecommand \@ifx [1]{%
 \ifx #1\expandafter \@firstoftwo
 \else \expandafter \@secondoftwo
 \fi
}%
\providecommand \natexlab [1]{#1}%
\providecommand \enquote  [1]{``#1''}%
\providecommand \bibnamefont  [1]{#1}%
\providecommand \bibfnamefont [1]{#1}%
\providecommand \citenamefont [1]{#1}%
\providecommand \href@noop [0]{\@secondoftwo}%
\providecommand \href [0]{\begingroup \@sanitize@url \@href}%
\providecommand \@href[1]{\@@startlink{#1}\@@href}%
\providecommand \@@href[1]{\endgroup#1\@@endlink}%
\providecommand \@sanitize@url [0]{\catcode `\\12\catcode `\$12\catcode
  `\&12\catcode `\#12\catcode `\^12\catcode `\_12\catcode `\%12\relax}%
\providecommand \@@startlink[1]{}%
\providecommand \@@endlink[0]{}%
\providecommand \url  [0]{\begingroup\@sanitize@url \@url }%
\providecommand \@url [1]{\endgroup\@href {#1}{\urlprefix }}%
\providecommand \urlprefix  [0]{URL }%
\providecommand \Eprint [0]{\href }%
\providecommand \doibase [0]{http://dx.doi.org/}%
\providecommand \selectlanguage [0]{\@gobble}%
\providecommand \bibinfo  [0]{\@secondoftwo}%
\providecommand \bibfield  [0]{\@secondoftwo}%
\providecommand \translation [1]{[#1]}%
\providecommand \BibitemOpen [0]{}%
\providecommand \bibitemStop [0]{}%
\providecommand \bibitemNoStop [0]{.\EOS\space}%
\providecommand \EOS [0]{\spacefactor3000\relax}%
\providecommand \BibitemShut  [1]{\csname bibitem#1\endcsname}%
\let\auto@bib@innerbib\@empty
\bibitem [{\citenamefont {Peruzzo}\ \emph {et~al.}(2014)\citenamefont
  {Peruzzo}, \citenamefont {McClean}, \citenamefont {Shadbolt}, \citenamefont
  {Yung}, \citenamefont {Zhou}, \citenamefont {Love}, \citenamefont
  {Aspuru-Guzik},\ and\ \citenamefont {O'brien}}]{peruzzo2014VQE}%
  \BibitemOpen
  \bibfield  {author} {\bibinfo {author} {\bibfnamefont {Alberto}\ \bibnamefont
  {Peruzzo}}, \bibinfo {author} {\bibfnamefont {Jarrod}\ \bibnamefont
  {McClean}}, \bibinfo {author} {\bibfnamefont {Peter}\ \bibnamefont
  {Shadbolt}}, \bibinfo {author} {\bibfnamefont {Man-Hong}\ \bibnamefont
  {Yung}}, \bibinfo {author} {\bibfnamefont {Xiao-Qi}\ \bibnamefont {Zhou}},
  \bibinfo {author} {\bibfnamefont {Peter~J}\ \bibnamefont {Love}}, \bibinfo
  {author} {\bibfnamefont {Al{\'a}n}\ \bibnamefont {Aspuru-Guzik}}, \ and\
  \bibinfo {author} {\bibfnamefont {Jeremy~L}\ \bibnamefont {O'brien}},\
  }\bibfield  {title} {\enquote {\bibinfo {title} {A variational eigenvalue
  solver on a photonic quantum processor},}\ }\href {\doibase
  10.1038/ncomms5213} {\bibfield  {journal} {\bibinfo  {journal} {Nature
  Communications}\ }\textbf {\bibinfo {volume} {5}},\ \bibinfo {pages} {4213}
  (\bibinfo {year} {2014})}\BibitemShut {NoStop}%
\bibitem [{\citenamefont {Farhi}\ \emph {et~al.}(2014)\citenamefont {Farhi},
  \citenamefont {Goldstone},\ and\ \citenamefont {Gutmann}}]{farhi2014QAOA}%
  \BibitemOpen
  \bibfield  {author} {\bibinfo {author} {\bibfnamefont {Edward}\ \bibnamefont
  {Farhi}}, \bibinfo {author} {\bibfnamefont {Jeffrey}\ \bibnamefont
  {Goldstone}}, \ and\ \bibinfo {author} {\bibfnamefont {Sam}\ \bibnamefont
  {Gutmann}},\ }\bibfield  {title} {\enquote {\bibinfo {title} {A quantum
  approximate optimization algorithm},}\ }\href
  {https://arxiv.org/abs/1411.4028} {\bibfield  {journal} {\bibinfo  {journal}
  {arXiv:1411.4028}\ } (\bibinfo {year} {2014})}\BibitemShut {NoStop}%
\bibitem [{\citenamefont {Johnson}\ \emph {et~al.}(2017)\citenamefont
  {Johnson}, \citenamefont {Romero}, \citenamefont {Olson}, \citenamefont
  {Cao},\ and\ \citenamefont {Aspuru-Guzik}}]{johnson2017qvector}%
  \BibitemOpen
  \bibfield  {author} {\bibinfo {author} {\bibfnamefont {Peter~D}\ \bibnamefont
  {Johnson}}, \bibinfo {author} {\bibfnamefont {Jonathan}\ \bibnamefont
  {Romero}}, \bibinfo {author} {\bibfnamefont {Jonathan}\ \bibnamefont
  {Olson}}, \bibinfo {author} {\bibfnamefont {Yudong}\ \bibnamefont {Cao}}, \
  and\ \bibinfo {author} {\bibfnamefont {Al{\'a}n}\ \bibnamefont
  {Aspuru-Guzik}},\ }\bibfield  {title} {\enquote {\bibinfo {title} {{QVECTOR}:
  an algorithm for device-tailored quantum error correction},}\ }\href
  {https://arxiv.org/abs/1711.02249} {\bibfield  {journal} {\bibinfo  {journal}
  {arXiv:1711.02249}\ } (\bibinfo {year} {2017})}\BibitemShut {NoStop}%
\bibitem [{\citenamefont {Romero}\ \emph {et~al.}(2017)\citenamefont {Romero},
  \citenamefont {Olson},\ and\ \citenamefont
  {Aspuru-Guzik}}]{romero2017quantum}%
  \BibitemOpen
  \bibfield  {author} {\bibinfo {author} {\bibfnamefont {Jonathan}\
  \bibnamefont {Romero}}, \bibinfo {author} {\bibfnamefont {Jonathan~P}\
  \bibnamefont {Olson}}, \ and\ \bibinfo {author} {\bibfnamefont {Alan}\
  \bibnamefont {Aspuru-Guzik}},\ }\bibfield  {title} {\enquote {\bibinfo
  {title} {Quantum autoencoders for efficient compression of quantum data},}\
  }\href {https://iopscience.iop.org/article/10.1088/2058-9565/aa8072}
  {\bibfield  {journal} {\bibinfo  {journal} {Quantum Science and Technology}\
  }\textbf {\bibinfo {volume} {2}},\ \bibinfo {pages} {045001} (\bibinfo {year}
  {2017})}\BibitemShut {NoStop}%
\bibitem [{\citenamefont {LaRose}\ \emph {et~al.}(2019)\citenamefont {LaRose},
  \citenamefont {Tikku}, \citenamefont {O'Neel-Judy}, \citenamefont {Cincio},\
  and\ \citenamefont {Coles}}]{larose2018}%
  \BibitemOpen
  \bibfield  {author} {\bibinfo {author} {\bibfnamefont {Ryan}\ \bibnamefont
  {LaRose}}, \bibinfo {author} {\bibfnamefont {Arkin}\ \bibnamefont {Tikku}},
  \bibinfo {author} {\bibfnamefont {{\'E}tude}\ \bibnamefont {O'Neel-Judy}},
  \bibinfo {author} {\bibfnamefont {Lukasz}\ \bibnamefont {Cincio}}, \ and\
  \bibinfo {author} {\bibfnamefont {Patrick~J}\ \bibnamefont {Coles}},\
  }\bibfield  {title} {\enquote {\bibinfo {title} {Variational quantum state
  diagonalization},}\ }\href
  {https://www.nature.com/articles/s41534-019-0167-6} {\bibfield  {journal}
  {\bibinfo  {journal} {npj Quantum Information}\ }\textbf {\bibinfo {volume}
  {5}},\ \bibinfo {pages} {57} (\bibinfo {year} {2019})}\BibitemShut {NoStop}%
\bibitem [{\citenamefont {Arrasmith}\ \emph {et~al.}(2019)\citenamefont
  {Arrasmith}, \citenamefont {Cincio}, \citenamefont {Sornborger},
  \citenamefont {Zurek},\ and\ \citenamefont
  {Coles}}]{arrasmith2019variational}%
  \BibitemOpen
  \bibfield  {author} {\bibinfo {author} {\bibfnamefont {Andrew}\ \bibnamefont
  {Arrasmith}}, \bibinfo {author} {\bibfnamefont {Lukasz}\ \bibnamefont
  {Cincio}}, \bibinfo {author} {\bibfnamefont {Andrew~T}\ \bibnamefont
  {Sornborger}}, \bibinfo {author} {\bibfnamefont {Wojciech~H}\ \bibnamefont
  {Zurek}}, \ and\ \bibinfo {author} {\bibfnamefont {Patrick~J}\ \bibnamefont
  {Coles}},\ }\bibfield  {title} {\enquote {\bibinfo {title} {Variational
  consistent histories as a hybrid algorithm for quantum foundations},}\ }\href
  {https://www.nature.com/articles/s41467-019-11417-0} {\bibfield  {journal}
  {\bibinfo  {journal} {Nature Communications}\ }\textbf {\bibinfo {volume}
  {10}},\ \bibinfo {pages} {3438} (\bibinfo {year} {2019})}\BibitemShut
  {NoStop}%
\bibitem [{\citenamefont {Cerezo}\ \emph {et~al.}(2019)\citenamefont {Cerezo},
  \citenamefont {Poremba}, \citenamefont {Cincio},\ and\ \citenamefont
  {Coles}}]{cerezo2019variational}%
  \BibitemOpen
  \bibfield  {author} {\bibinfo {author} {\bibfnamefont {M}~\bibnamefont
  {Cerezo}}, \bibinfo {author} {\bibfnamefont {Alexander}\ \bibnamefont
  {Poremba}}, \bibinfo {author} {\bibfnamefont {Lukasz}\ \bibnamefont
  {Cincio}}, \ and\ \bibinfo {author} {\bibfnamefont {Patrick~J}\ \bibnamefont
  {Coles}},\ }\bibfield  {title} {\enquote {\bibinfo {title} {Variational
  quantum fidelity estimation},}\ }\href {https://arxiv.org/abs/1906.09253}
  {\bibfield  {journal} {\bibinfo  {journal} {arXiv:1906.09253}\ } (\bibinfo
  {year} {2019})}\BibitemShut {NoStop}%
\bibitem [{\citenamefont {Jones}\ \emph {et~al.}(2019)\citenamefont {Jones},
  \citenamefont {Endo}, \citenamefont {McArdle}, \citenamefont {Yuan},\ and\
  \citenamefont {Benjamin}}]{jones2019variational}%
  \BibitemOpen
  \bibfield  {author} {\bibinfo {author} {\bibfnamefont {Tyson}\ \bibnamefont
  {Jones}}, \bibinfo {author} {\bibfnamefont {Suguru}\ \bibnamefont {Endo}},
  \bibinfo {author} {\bibfnamefont {Sam}\ \bibnamefont {McArdle}}, \bibinfo
  {author} {\bibfnamefont {Xiao}\ \bibnamefont {Yuan}}, \ and\ \bibinfo
  {author} {\bibfnamefont {Simon~C}\ \bibnamefont {Benjamin}},\ }\bibfield
  {title} {\enquote {\bibinfo {title} {Variational quantum algorithms for
  discovering hamiltonian spectra},}\ }\href
  {https://journals.aps.org/pra/abstract/10.1103/PhysRevA.99.062304} {\bibfield
   {journal} {\bibinfo  {journal} {Physical Review A}\ }\textbf {\bibinfo
  {volume} {99}},\ \bibinfo {pages} {062304} (\bibinfo {year}
  {2019})}\BibitemShut {NoStop}%
\bibitem [{\citenamefont {Yuan}\ \emph {et~al.}(2018)\citenamefont {Yuan},
  \citenamefont {Endo}, \citenamefont {Zhao}, \citenamefont {Benjamin},\ and\
  \citenamefont {Li}}]{yuan2018theory}%
  \BibitemOpen
  \bibfield  {author} {\bibinfo {author} {\bibfnamefont {Xiao}\ \bibnamefont
  {Yuan}}, \bibinfo {author} {\bibfnamefont {Suguru}\ \bibnamefont {Endo}},
  \bibinfo {author} {\bibfnamefont {Qi}~\bibnamefont {Zhao}}, \bibinfo {author}
  {\bibfnamefont {Simon}\ \bibnamefont {Benjamin}}, \ and\ \bibinfo {author}
  {\bibfnamefont {Ying}\ \bibnamefont {Li}},\ }\bibfield  {title} {\enquote
  {\bibinfo {title} {Theory of variational quantum simulation},}\ }\href
  {https://arxiv.org/abs/1812.08767} {\bibfield  {journal} {\bibinfo  {journal}
  {arXiv:1812.08767}\ } (\bibinfo {year} {2018})}\BibitemShut {NoStop}%
\bibitem [{\citenamefont {Li}\ and\ \citenamefont
  {Benjamin}(2017)}]{li2017efficient}%
  \BibitemOpen
  \bibfield  {author} {\bibinfo {author} {\bibfnamefont {Ying}\ \bibnamefont
  {Li}}\ and\ \bibinfo {author} {\bibfnamefont {Simon~C}\ \bibnamefont
  {Benjamin}},\ }\bibfield  {title} {\enquote {\bibinfo {title} {Efficient
  variational quantum simulator incorporating active error minimization},}\
  }\href {https://journals.aps.org/prx/abstract/10.1103/PhysRevX.7.021050}
  {\bibfield  {journal} {\bibinfo  {journal} {Physical Review X}\ }\textbf
  {\bibinfo {volume} {7}},\ \bibinfo {pages} {021050} (\bibinfo {year}
  {2017})}\BibitemShut {NoStop}%
\bibitem [{\citenamefont {Kokail}\ \emph {et~al.}(2019)\citenamefont {Kokail},
  \citenamefont {Maier}, \citenamefont {van Bijnen}, \citenamefont {Brydges},
  \citenamefont {Joshi}, \citenamefont {Jurcevic}, \citenamefont {Muschik},
  \citenamefont {Silvi}, \citenamefont {Blatt}, \citenamefont {Roos} \emph
  {et~al.}}]{kokail2019self}%
  \BibitemOpen
  \bibfield  {author} {\bibinfo {author} {\bibfnamefont {C}~\bibnamefont
  {Kokail}}, \bibinfo {author} {\bibfnamefont {C}~\bibnamefont {Maier}},
  \bibinfo {author} {\bibfnamefont {R}~\bibnamefont {van Bijnen}}, \bibinfo
  {author} {\bibfnamefont {T}~\bibnamefont {Brydges}}, \bibinfo {author}
  {\bibfnamefont {MK}~\bibnamefont {Joshi}}, \bibinfo {author} {\bibfnamefont
  {P}~\bibnamefont {Jurcevic}}, \bibinfo {author} {\bibfnamefont
  {CA}~\bibnamefont {Muschik}}, \bibinfo {author} {\bibfnamefont
  {P}~\bibnamefont {Silvi}}, \bibinfo {author} {\bibfnamefont {R}~\bibnamefont
  {Blatt}}, \bibinfo {author} {\bibfnamefont {CF}~\bibnamefont {Roos}},  \emph
  {et~al.},\ }\bibfield  {title} {\enquote {\bibinfo {title} {Self-verifying
  variational quantum simulation of lattice models},}\ }\href
  {https://www.nature.com/articles/s41586-019-1177-4} {\bibfield  {journal}
  {\bibinfo  {journal} {Nature}\ }\textbf {\bibinfo {volume} {569}},\ \bibinfo
  {pages} {355} (\bibinfo {year} {2019})}\BibitemShut {NoStop}%
\bibitem [{\citenamefont {Khatri}\ \emph {et~al.}(2019)\citenamefont {Khatri},
  \citenamefont {LaRose}, \citenamefont {Poremba}, \citenamefont {Cincio},
  \citenamefont {Sornborger},\ and\ \citenamefont
  {Coles}}]{Khatri2019quantumassisted}%
  \BibitemOpen
  \bibfield  {author} {\bibinfo {author} {\bibfnamefont {Sumeet}\ \bibnamefont
  {Khatri}}, \bibinfo {author} {\bibfnamefont {Ryan}\ \bibnamefont {LaRose}},
  \bibinfo {author} {\bibfnamefont {Alexander}\ \bibnamefont {Poremba}},
  \bibinfo {author} {\bibfnamefont {Lukasz}\ \bibnamefont {Cincio}}, \bibinfo
  {author} {\bibfnamefont {Andrew~T}\ \bibnamefont {Sornborger}}, \ and\
  \bibinfo {author} {\bibfnamefont {Patrick~J}\ \bibnamefont {Coles}},\
  }\bibfield  {title} {\enquote {\bibinfo {title} {Quantum-assisted quantum
  compiling},}\ }\href {\doibase 10.22331/q-2019-05-13-140} {\bibfield
  {journal} {\bibinfo  {journal} {{Quantum}}\ }\textbf {\bibinfo {volume}
  {3}},\ \bibinfo {pages} {140} (\bibinfo {year} {2019})}\BibitemShut {NoStop}%
\bibitem [{\citenamefont {Jones}\ and\ \citenamefont
  {Benjamin}(2018)}]{jones2018quantum}%
  \BibitemOpen
  \bibfield  {author} {\bibinfo {author} {\bibfnamefont {Tyson}\ \bibnamefont
  {Jones}}\ and\ \bibinfo {author} {\bibfnamefont {Simon~C}\ \bibnamefont
  {Benjamin}},\ }\bibfield  {title} {\enquote {\bibinfo {title} {Quantum
  compilation and circuit optimisation via energy dissipation},}\ }\href
  {https://arxiv.org/abs/1811.03147} {\bibfield  {journal} {\bibinfo  {journal}
  {arXiv:1811.03147}\ } (\bibinfo {year} {2018})}\BibitemShut {NoStop}%
\bibitem [{\citenamefont {Heya}\ \emph {et~al.}(2018)\citenamefont {Heya},
  \citenamefont {Suzuki}, \citenamefont {Nakamura},\ and\ \citenamefont
  {Fujii}}]{heya2018variational}%
  \BibitemOpen
  \bibfield  {author} {\bibinfo {author} {\bibfnamefont {Kentaro}\ \bibnamefont
  {Heya}}, \bibinfo {author} {\bibfnamefont {Yasunari}\ \bibnamefont {Suzuki}},
  \bibinfo {author} {\bibfnamefont {Yasunobu}\ \bibnamefont {Nakamura}}, \ and\
  \bibinfo {author} {\bibfnamefont {Keisuke}\ \bibnamefont {Fujii}},\
  }\bibfield  {title} {\enquote {\bibinfo {title} {Variational quantum gate
  optimization},}\ }\href {https://arxiv.org/abs/1810.12745} {\bibfield
  {journal} {\bibinfo  {journal} {arXiv:1810.12745}\ } (\bibinfo {year}
  {2018})}\BibitemShut {NoStop}%
\bibitem [{\citenamefont {Endo}\ \emph {et~al.}(2018)\citenamefont {Endo},
  \citenamefont {Li}, \citenamefont {Benjamin},\ and\ \citenamefont
  {Yuan}}]{endo2018variational}%
  \BibitemOpen
  \bibfield  {author} {\bibinfo {author} {\bibfnamefont {Suguru}\ \bibnamefont
  {Endo}}, \bibinfo {author} {\bibfnamefont {Ying}\ \bibnamefont {Li}},
  \bibinfo {author} {\bibfnamefont {Simon}\ \bibnamefont {Benjamin}}, \ and\
  \bibinfo {author} {\bibfnamefont {Xiao}\ \bibnamefont {Yuan}},\ }\bibfield
  {title} {\enquote {\bibinfo {title} {Variational quantum simulation of
  general processes},}\ }\href@noop {} {\bibfield  {journal} {\bibinfo
  {journal} {arXiv preprint arXiv:1812.08778}\ } (\bibinfo {year}
  {2018})}\BibitemShut {NoStop}%
\bibitem [{\citenamefont {Sharma}\ \emph {et~al.}(2019)\citenamefont {Sharma},
  \citenamefont {Khatri}, \citenamefont {Cerezo},\ and\ \citenamefont
  {Coles}}]{sharma2019noise}%
  \BibitemOpen
  \bibfield  {author} {\bibinfo {author} {\bibfnamefont {Kunal}\ \bibnamefont
  {Sharma}}, \bibinfo {author} {\bibfnamefont {Sumeet}\ \bibnamefont {Khatri}},
  \bibinfo {author} {\bibfnamefont {M}~\bibnamefont {Cerezo}}, \ and\ \bibinfo
  {author} {\bibfnamefont {Patrick~J}\ \bibnamefont {Coles}},\ }\bibfield
  {title} {\enquote {\bibinfo {title} {Noise resilience of variational quantum
  compiling},}\ }\href {https://arxiv.org/abs/1908.04416} {\bibfield  {journal}
  {\bibinfo  {journal} {arXiv:1908.04416}\ } (\bibinfo {year}
  {2019})}\BibitemShut {NoStop}%
\bibitem [{\citenamefont {Carolan}\ \emph {et~al.}(2019)\citenamefont
  {Carolan}, \citenamefont {Mosheni}, \citenamefont {Olson}, \citenamefont
  {Prabhu}, \citenamefont {Chen}, \citenamefont {Bunandar}, \citenamefont
  {Harris}, \citenamefont {Wong}, \citenamefont {Hochberg}, \citenamefont
  {Lloyd} \emph {et~al.}}]{carolan2019variational}%
  \BibitemOpen
  \bibfield  {author} {\bibinfo {author} {\bibfnamefont {Jacques}\ \bibnamefont
  {Carolan}}, \bibinfo {author} {\bibfnamefont {Masoud}\ \bibnamefont
  {Mosheni}}, \bibinfo {author} {\bibfnamefont {Jonathan~P}\ \bibnamefont
  {Olson}}, \bibinfo {author} {\bibfnamefont {Mihika}\ \bibnamefont {Prabhu}},
  \bibinfo {author} {\bibfnamefont {Changchen}\ \bibnamefont {Chen}}, \bibinfo
  {author} {\bibfnamefont {Darius}\ \bibnamefont {Bunandar}}, \bibinfo {author}
  {\bibfnamefont {Nicholas~C}\ \bibnamefont {Harris}}, \bibinfo {author}
  {\bibfnamefont {Franco~NC}\ \bibnamefont {Wong}}, \bibinfo {author}
  {\bibfnamefont {Michael}\ \bibnamefont {Hochberg}}, \bibinfo {author}
  {\bibfnamefont {Seth}\ \bibnamefont {Lloyd}},  \emph {et~al.},\ }\bibfield
  {title} {\enquote {\bibinfo {title} {Variational quantum unsampling on a
  quantum photonic processor},}\ }\href {https://arxiv.org/abs/1904.10463}
  {\bibfield  {journal} {\bibinfo  {journal} {arXiv:1904.10463}\ } (\bibinfo
  {year} {2019})}\BibitemShut {NoStop}%
\bibitem [{\citenamefont {Yoshioka}\ \emph {et~al.}(2019)\citenamefont
  {Yoshioka}, \citenamefont {Nakagawa}, \citenamefont {Mitarai},\ and\
  \citenamefont {Fujii}}]{yoshioka2019variational}%
  \BibitemOpen
  \bibfield  {author} {\bibinfo {author} {\bibfnamefont {Nobuyuki}\
  \bibnamefont {Yoshioka}}, \bibinfo {author} {\bibfnamefont {Yuya~O}\
  \bibnamefont {Nakagawa}}, \bibinfo {author} {\bibfnamefont {Kosuke}\
  \bibnamefont {Mitarai}}, \ and\ \bibinfo {author} {\bibfnamefont {Keisuke}\
  \bibnamefont {Fujii}},\ }\bibfield  {title} {\enquote {\bibinfo {title}
  {Variational quantum algorithm for non-equilirium steady states},}\ }\href
  {https://arxiv.org/abs/1908.09836} {\bibfield  {journal} {\bibinfo  {journal}
  {arXiv:1908.09836}\ } (\bibinfo {year} {2019})}\BibitemShut {NoStop}%
\bibitem [{\citenamefont {Bravo-Prieto}\ \emph
  {et~al.}(2019{\natexlab{a}})\citenamefont {Bravo-Prieto}, \citenamefont
  {LaRose}, \citenamefont {Cerezo}, \citenamefont {Subasi}, \citenamefont
  {Cincio},\ and\ \citenamefont {Coles}}]{bravo-prieto2019}%
  \BibitemOpen
  \bibfield  {author} {\bibinfo {author} {\bibfnamefont {Carlos}\ \bibnamefont
  {Bravo-Prieto}}, \bibinfo {author} {\bibnamefont {LaRose}}, \bibinfo {author}
  {\bibfnamefont {M.}~\bibnamefont {Cerezo}}, \bibinfo {author} {\bibfnamefont
  {Yigit}\ \bibnamefont {Subasi}}, \bibinfo {author} {\bibfnamefont {Lukasz}\
  \bibnamefont {Cincio}}, \ and\ \bibinfo {author} {\bibfnamefont {Patrick~J.}\
  \bibnamefont {Coles}},\ }\bibfield  {title} {\enquote {\bibinfo {title}
  {Variational quantum linear solver: A hybrid algorithm for linear systems},}\
  }\href {https://arxiv.org/abs/1909.05820} {\bibfield  {journal} {\bibinfo
  {journal} {arXiv:1909.05820}\ } (\bibinfo {year}
  {2019}{\natexlab{a}})}\BibitemShut {NoStop}%
\bibitem [{\citenamefont {Xu}\ \emph {et~al.}(2019)\citenamefont {Xu},
  \citenamefont {Sun}, \citenamefont {Endo}, \citenamefont {Li}, \citenamefont
  {Benjamin},\ and\ \citenamefont {Yuan}}]{xu2019variational}%
  \BibitemOpen
  \bibfield  {author} {\bibinfo {author} {\bibfnamefont {Xiaosi}\ \bibnamefont
  {Xu}}, \bibinfo {author} {\bibfnamefont {Jinzhao}\ \bibnamefont {Sun}},
  \bibinfo {author} {\bibfnamefont {Suguru}\ \bibnamefont {Endo}}, \bibinfo
  {author} {\bibfnamefont {Ying}\ \bibnamefont {Li}}, \bibinfo {author}
  {\bibfnamefont {Simon~C}\ \bibnamefont {Benjamin}}, \ and\ \bibinfo {author}
  {\bibfnamefont {Xiao}\ \bibnamefont {Yuan}},\ }\bibfield  {title} {\enquote
  {\bibinfo {title} {Variational algorithms for linear algebra},}\ }\href
  {https://arxiv.org/abs/1909.03898} {\bibfield  {journal} {\bibinfo  {journal}
  {arXiv preprint arXiv:1909.03898}\ } (\bibinfo {year} {2019})}\BibitemShut
  {NoStop}%
\bibitem [{\citenamefont {McArdle}\ \emph {et~al.}(2019)\citenamefont
  {McArdle}, \citenamefont {Jones}, \citenamefont {Endo}, \citenamefont {Li},
  \citenamefont {Benjamin},\ and\ \citenamefont
  {Yuan}}]{mcardle2019variational}%
  \BibitemOpen
  \bibfield  {author} {\bibinfo {author} {\bibfnamefont {Sam}\ \bibnamefont
  {McArdle}}, \bibinfo {author} {\bibfnamefont {Tyson}\ \bibnamefont {Jones}},
  \bibinfo {author} {\bibfnamefont {Suguru}\ \bibnamefont {Endo}}, \bibinfo
  {author} {\bibfnamefont {Ying}\ \bibnamefont {Li}}, \bibinfo {author}
  {\bibfnamefont {Simon~C}\ \bibnamefont {Benjamin}}, \ and\ \bibinfo {author}
  {\bibfnamefont {Xiao}\ \bibnamefont {Yuan}},\ }\bibfield  {title} {\enquote
  {\bibinfo {title} {Variational ansatz-based quantum simulation of imaginary
  time evolution},}\ }\href {https://www.nature.com/articles/s41534-019-0187-2}
  {\bibfield  {journal} {\bibinfo  {journal} {npj Quantum Information}\
  }\textbf {\bibinfo {volume} {5}},\ \bibinfo {pages} {1--6} (\bibinfo {year}
  {2019})}\BibitemShut {NoStop}%
\bibitem [{\citenamefont {Cirstoiu}\ \emph {et~al.}(2019)\citenamefont
  {Cirstoiu}, \citenamefont {Holmes}, \citenamefont {Iosue}, \citenamefont
  {Cincio}, \citenamefont {Coles},\ and\ \citenamefont
  {Sornborger}}]{cirstoiu2019variational}%
  \BibitemOpen
  \bibfield  {author} {\bibinfo {author} {\bibfnamefont {Cristina}\
  \bibnamefont {Cirstoiu}}, \bibinfo {author} {\bibfnamefont {Zoe}\
  \bibnamefont {Holmes}}, \bibinfo {author} {\bibfnamefont {Joseph}\
  \bibnamefont {Iosue}}, \bibinfo {author} {\bibfnamefont {Lukasz}\
  \bibnamefont {Cincio}}, \bibinfo {author} {\bibfnamefont {Patrick~J}\
  \bibnamefont {Coles}}, \ and\ \bibinfo {author} {\bibfnamefont {Andrew}\
  \bibnamefont {Sornborger}},\ }\bibfield  {title} {\enquote {\bibinfo {title}
  {Variational fast forwarding for quantum simulation beyond the coherence
  time},}\ }\href {https://arxiv.org/abs/1910.04292} {\bibfield  {journal}
  {\bibinfo  {journal} {arXiv preprint arXiv:1910.04292}\ } (\bibinfo {year}
  {2019})}\BibitemShut {NoStop}%
\bibitem [{\citenamefont {Otten}\ \emph {et~al.}(2019)\citenamefont {Otten},
  \citenamefont {Cortes},\ and\ \citenamefont {Gray}}]{otten2019noise}%
  \BibitemOpen
  \bibfield  {author} {\bibinfo {author} {\bibfnamefont {Matthew}\ \bibnamefont
  {Otten}}, \bibinfo {author} {\bibfnamefont {Cristian~L}\ \bibnamefont
  {Cortes}}, \ and\ \bibinfo {author} {\bibfnamefont {Stephen~K}\ \bibnamefont
  {Gray}},\ }\bibfield  {title} {\enquote {\bibinfo {title} {Noise-resilient
  quantum dynamics using symmetry-preserving ansatzes},}\ }\href
  {https://arxiv.org/abs/1910.06284} {\bibfield  {journal} {\bibinfo  {journal}
  {arXiv preprint arXiv:1910.06284}\ } (\bibinfo {year} {2019})}\BibitemShut
  {NoStop}%
\bibitem [{\citenamefont {Lubasch}\ \emph {et~al.}(2020)\citenamefont
  {Lubasch}, \citenamefont {Joo}, \citenamefont {Moinier}, \citenamefont
  {Kiffner},\ and\ \citenamefont {Jaksch}}]{LubaschVariational20}%
  \BibitemOpen
  \bibfield  {author} {\bibinfo {author} {\bibfnamefont {Michael}\ \bibnamefont
  {Lubasch}}, \bibinfo {author} {\bibfnamefont {Jaewoo}\ \bibnamefont {Joo}},
  \bibinfo {author} {\bibfnamefont {Pierre}\ \bibnamefont {Moinier}}, \bibinfo
  {author} {\bibfnamefont {Martin}\ \bibnamefont {Kiffner}}, \ and\ \bibinfo
  {author} {\bibfnamefont {Dieter}\ \bibnamefont {Jaksch}},\ }\bibfield
  {title} {\enquote {\bibinfo {title} {Variational quantum algorithms for
  nonlinear problems},}\ }\href {\doibase 10.1103/PhysRevA.101.010301}
  {\bibfield  {journal} {\bibinfo  {journal} {Phys. Rev. A}\ }\textbf {\bibinfo
  {volume} {101}},\ \bibinfo {pages} {010301} (\bibinfo {year}
  {2020})}\BibitemShut {NoStop}%
\bibitem [{\citenamefont {Verdon}\ \emph
  {et~al.}(2019{\natexlab{a}})\citenamefont {Verdon}, \citenamefont {Marks},
  \citenamefont {Nanda}, \citenamefont {Leichenauer},\ and\ \citenamefont
  {Hidary}}]{verdon2019quantum}%
  \BibitemOpen
  \bibfield  {author} {\bibinfo {author} {\bibfnamefont {Guillaume}\
  \bibnamefont {Verdon}}, \bibinfo {author} {\bibfnamefont {Jacob}\
  \bibnamefont {Marks}}, \bibinfo {author} {\bibfnamefont {Sasha}\ \bibnamefont
  {Nanda}}, \bibinfo {author} {\bibfnamefont {Stefan}\ \bibnamefont
  {Leichenauer}}, \ and\ \bibinfo {author} {\bibfnamefont {Jack}\ \bibnamefont
  {Hidary}},\ }\bibfield  {title} {\enquote {\bibinfo {title} {Quantum
  hamiltonian-based models and the variational quantum thermalizer
  algorithm},}\ }\href {https://arxiv.org/abs/1910.02071} {\bibfield  {journal}
  {\bibinfo  {journal} {arXiv preprint arXiv:1910.02071}\ } (\bibinfo {year}
  {2019}{\natexlab{a}})}\BibitemShut {NoStop}%
\bibitem [{\citenamefont {Bravo-Prieto}\ \emph
  {et~al.}(2019{\natexlab{b}})\citenamefont {Bravo-Prieto}, \citenamefont
  {Garc{\'\i}a-Mart{\'\i}n},\ and\ \citenamefont {Latorre}}]{bravo2019quantum}%
  \BibitemOpen
  \bibfield  {author} {\bibinfo {author} {\bibfnamefont {Carlos}\ \bibnamefont
  {Bravo-Prieto}}, \bibinfo {author} {\bibfnamefont {Diego}\ \bibnamefont
  {Garc{\'\i}a-Mart{\'\i}n}}, \ and\ \bibinfo {author} {\bibfnamefont
  {Jos{\'e}~I}\ \bibnamefont {Latorre}},\ }\bibfield  {title} {\enquote
  {\bibinfo {title} {Quantum singular value decomposer},}\ }\href
  {https://arxiv.org/abs/1905.01353} {\bibfield  {journal} {\bibinfo  {journal}
  {arXiv preprint arXiv:1905.01353}\ } (\bibinfo {year}
  {2019}{\natexlab{b}})}\BibitemShut {NoStop}%
\bibitem [{\citenamefont {Cerezo}\ \emph
  {et~al.}(2020{\natexlab{a}})\citenamefont {Cerezo}, \citenamefont {Sharma},
  \citenamefont {Arrasmith},\ and\ \citenamefont
  {Coles}}]{cerezo2020variational}%
  \BibitemOpen
  \bibfield  {author} {\bibinfo {author} {\bibfnamefont {M}~\bibnamefont
  {Cerezo}}, \bibinfo {author} {\bibfnamefont {Kunal}\ \bibnamefont {Sharma}},
  \bibinfo {author} {\bibfnamefont {Andrew}\ \bibnamefont {Arrasmith}}, \ and\
  \bibinfo {author} {\bibfnamefont {Patrick~J}\ \bibnamefont {Coles}},\
  }\bibfield  {title} {\enquote {\bibinfo {title} {Variational quantum state
  eigensolver},}\ }\href {https://arxiv.org/abs/2004.01372} {\bibfield
  {journal} {\bibinfo  {journal} {arXiv preprint arXiv:2004.01372}\ } (\bibinfo
  {year} {2020}{\natexlab{a}})}\BibitemShut {NoStop}%
\bibitem [{\citenamefont {Mitarai}\ \emph {et~al.}(2018)\citenamefont
  {Mitarai}, \citenamefont {Negoro}, \citenamefont {Kitagawa},\ and\
  \citenamefont {Fujii}}]{mitarai2018quantum}%
  \BibitemOpen
  \bibfield  {author} {\bibinfo {author} {\bibfnamefont {Kosuke}\ \bibnamefont
  {Mitarai}}, \bibinfo {author} {\bibfnamefont {Makoto}\ \bibnamefont
  {Negoro}}, \bibinfo {author} {\bibfnamefont {Masahiro}\ \bibnamefont
  {Kitagawa}}, \ and\ \bibinfo {author} {\bibfnamefont {Keisuke}\ \bibnamefont
  {Fujii}},\ }\bibfield  {title} {\enquote {\bibinfo {title} {Quantum circuit
  learning},}\ }\href {\doibase 10.1103/PhysRevA.98.032309} {\bibfield
  {journal} {\bibinfo  {journal} {Physical Review A}\ }\textbf {\bibinfo
  {volume} {98}},\ \bibinfo {eid} {032309} (\bibinfo {year}
  {2018})}\BibitemShut {NoStop}%
\bibitem [{\citenamefont {Schuld}\ \emph {et~al.}(2019)\citenamefont {Schuld},
  \citenamefont {Bergholm}, \citenamefont {Gogolin}, \citenamefont {Izaac},\
  and\ \citenamefont {Killoran}}]{Schuld2019}%
  \BibitemOpen
  \bibfield  {author} {\bibinfo {author} {\bibfnamefont {Maria}\ \bibnamefont
  {Schuld}}, \bibinfo {author} {\bibfnamefont {Ville}\ \bibnamefont
  {Bergholm}}, \bibinfo {author} {\bibfnamefont {Christian}\ \bibnamefont
  {Gogolin}}, \bibinfo {author} {\bibfnamefont {Josh}\ \bibnamefont {Izaac}}, \
  and\ \bibinfo {author} {\bibfnamefont {Nathan}\ \bibnamefont {Killoran}},\
  }\bibfield  {title} {\enquote {\bibinfo {title} {Evaluating analytic
  gradients on quantum hardware},}\ }\href {\doibase
  10.1103/PhysRevA.99.032331} {\bibfield  {journal} {\bibinfo  {journal} {Phys.
  Rev. A}\ }\textbf {\bibinfo {volume} {99}},\ \bibinfo {pages} {032331}
  (\bibinfo {year} {2019})}\BibitemShut {NoStop}%
\bibitem [{\citenamefont {McClean}\ \emph {et~al.}(2018)\citenamefont
  {McClean}, \citenamefont {Boixo}, \citenamefont {Smelyanskiy}, \citenamefont
  {Babbush},\ and\ \citenamefont {Neven}}]{mcclean2018barren}%
  \BibitemOpen
  \bibfield  {author} {\bibinfo {author} {\bibfnamefont {J.~R.}\ \bibnamefont
  {McClean}}, \bibinfo {author} {\bibfnamefont {S.}~\bibnamefont {Boixo}},
  \bibinfo {author} {\bibfnamefont {V.~N.}\ \bibnamefont {Smelyanskiy}},
  \bibinfo {author} {\bibfnamefont {R.}~\bibnamefont {Babbush}}, \ and\
  \bibinfo {author} {\bibfnamefont {H.}~\bibnamefont {Neven}},\ }\bibfield
  {title} {\enquote {\bibinfo {title} {Barren plateaus in quantum neural
  network training landscapes},}\ }\href {\doibase 10.1038/s41467-018-07090-4}
  {\bibfield  {journal} {\bibinfo  {journal} {Nature Communications}\ }\textbf
  {\bibinfo {volume} {9}},\ \bibinfo {eid} {4812} (\bibinfo {year}
  {2018})}\BibitemShut {NoStop}%
\bibitem [{\citenamefont {Cerezo}\ \emph
  {et~al.}(2020{\natexlab{b}})\citenamefont {Cerezo}, \citenamefont {Sone},
  \citenamefont {Volkoff}, \citenamefont {Cincio},\ and\ \citenamefont
  {Coles}}]{cerezo2020cost}%
  \BibitemOpen
  \bibfield  {author} {\bibinfo {author} {\bibfnamefont {M}~\bibnamefont
  {Cerezo}}, \bibinfo {author} {\bibfnamefont {Akira}\ \bibnamefont {Sone}},
  \bibinfo {author} {\bibfnamefont {Tyler}\ \bibnamefont {Volkoff}}, \bibinfo
  {author} {\bibfnamefont {Lukasz}\ \bibnamefont {Cincio}}, \ and\ \bibinfo
  {author} {\bibfnamefont {Patrick~J}\ \bibnamefont {Coles}},\ }\bibfield
  {title} {\enquote {\bibinfo {title} {Cost-function-dependent barren plateaus
  in shallow quantum neural networks},}\ }\href@noop {} {\bibfield  {journal}
  {\bibinfo  {journal} {arXiv preprint arXiv:2001.00550}\ } (\bibinfo {year}
  {2020}{\natexlab{b}})}\BibitemShut {NoStop}%
\bibitem [{\citenamefont {McClean}\ \emph {et~al.}(2016)\citenamefont
  {McClean}, \citenamefont {Romero}, \citenamefont {Babbush},\ and\
  \citenamefont {Aspuru-Guzik}}]{mcclean2016theory}%
  \BibitemOpen
  \bibfield  {author} {\bibinfo {author} {\bibfnamefont {Jarrod~R}\
  \bibnamefont {McClean}}, \bibinfo {author} {\bibfnamefont {Jonathan}\
  \bibnamefont {Romero}}, \bibinfo {author} {\bibfnamefont {Ryan}\ \bibnamefont
  {Babbush}}, \ and\ \bibinfo {author} {\bibfnamefont {Al{\'a}n}\ \bibnamefont
  {Aspuru-Guzik}},\ }\bibfield  {title} {\enquote {\bibinfo {title} {The theory
  of variational hybrid quantum-classical algorithms},}\ }\href
  {https://iopscience.iop.org/article/10.1088/1367-2630/18/2/023023/meta}
  {\bibfield  {journal} {\bibinfo  {journal} {New Journal of Physics}\ }\textbf
  {\bibinfo {volume} {18}},\ \bibinfo {pages} {023023} (\bibinfo {year}
  {2016})}\BibitemShut {NoStop}%
\bibitem [{\citenamefont {Jena}\ \emph {et~al.}(2019)\citenamefont {Jena},
  \citenamefont {Genin},\ and\ \citenamefont {Mosca}}]{Jena2019}%
  \BibitemOpen
  \bibfield  {author} {\bibinfo {author} {\bibfnamefont {Andrew}\ \bibnamefont
  {Jena}}, \bibinfo {author} {\bibfnamefont {Scott}\ \bibnamefont {Genin}}, \
  and\ \bibinfo {author} {\bibfnamefont {Michele}\ \bibnamefont {Mosca}},\
  }\bibfield  {title} {\enquote {\bibinfo {title} {Pauli partitioning with
  respect to gate sets},}\ }\href {https://arxiv.org/abs/1907.07859} {\bibfield
   {journal} {\bibinfo  {journal} {arXiv:1907.07859}\ } (\bibinfo {year}
  {2019})}\BibitemShut {NoStop}%
\bibitem [{\citenamefont {Izmaylov}\ \emph {et~al.}(2019)\citenamefont
  {Izmaylov}, \citenamefont {Yen}, \citenamefont {Lang},\ and\ \citenamefont
  {Verteletskyi}}]{Izmaylov2019}%
  \BibitemOpen
  \bibfield  {author} {\bibinfo {author} {\bibfnamefont {Artur~F}\ \bibnamefont
  {Izmaylov}}, \bibinfo {author} {\bibfnamefont {Tzu-Ching}\ \bibnamefont
  {Yen}}, \bibinfo {author} {\bibfnamefont {Robert~A}\ \bibnamefont {Lang}}, \
  and\ \bibinfo {author} {\bibfnamefont {Vladyslav}\ \bibnamefont
  {Verteletskyi}},\ }\bibfield  {title} {\enquote {\bibinfo {title} {Unitary
  partitioning approach to the measurement problem in the variational quantum
  eigensolver method},}\ }\href {https://arxiv.org/abs/1907.09040} {\bibfield
  {journal} {\bibinfo  {journal} {arXiv:1907.09040}\ } (\bibinfo {year}
  {2019})}\BibitemShut {NoStop}%
\bibitem [{\citenamefont {Yen}\ \emph {et~al.}(2019)\citenamefont {Yen},
  \citenamefont {Verteletskyi},\ and\ \citenamefont {Izmaylov}}]{Yen2019}%
  \BibitemOpen
  \bibfield  {author} {\bibinfo {author} {\bibfnamefont {Tzu-Ching}\
  \bibnamefont {Yen}}, \bibinfo {author} {\bibfnamefont {Vladyslav}\
  \bibnamefont {Verteletskyi}}, \ and\ \bibinfo {author} {\bibfnamefont
  {Artur~F}\ \bibnamefont {Izmaylov}},\ }\bibfield  {title} {\enquote {\bibinfo
  {title} {Measuring all compatible operators in one series of a single-qubit
  measurements using unitary transformations},}\ }\href
  {https://arxiv.org/abs/1907.09386} {\bibfield  {journal} {\bibinfo  {journal}
  {arXiv:1907.09386}\ } (\bibinfo {year} {2019})}\BibitemShut {NoStop}%
\bibitem [{\citenamefont {Gokhale}\ \emph {et~al.}(2019)\citenamefont
  {Gokhale}, \citenamefont {Angiuli}, \citenamefont {Ding}, \citenamefont
  {Gui}, \citenamefont {Tomesh}, \citenamefont {Suchara}, \citenamefont
  {Martonosi},\ and\ \citenamefont {Chong}}]{Gokhale2019}%
  \BibitemOpen
  \bibfield  {author} {\bibinfo {author} {\bibfnamefont {Pranav}\ \bibnamefont
  {Gokhale}}, \bibinfo {author} {\bibfnamefont {Olivia}\ \bibnamefont
  {Angiuli}}, \bibinfo {author} {\bibfnamefont {Yongshan}\ \bibnamefont
  {Ding}}, \bibinfo {author} {\bibfnamefont {Kaiwen}\ \bibnamefont {Gui}},
  \bibinfo {author} {\bibfnamefont {Teague}\ \bibnamefont {Tomesh}}, \bibinfo
  {author} {\bibfnamefont {Martin}\ \bibnamefont {Suchara}}, \bibinfo {author}
  {\bibfnamefont {Margaret}\ \bibnamefont {Martonosi}}, \ and\ \bibinfo
  {author} {\bibfnamefont {Frederic~T}\ \bibnamefont {Chong}},\ }\bibfield
  {title} {\enquote {\bibinfo {title} {Minimizing state preparations in
  variational quantum eigensolver by partitioning into commuting families},}\
  }\href {https://arxiv.org/abs/1907.13623} {\bibfield  {journal} {\bibinfo
  {journal} {arXiv:1907.13623}\ } (\bibinfo {year} {2019})}\BibitemShut
  {NoStop}%
\bibitem [{\citenamefont {Crawford}\ \emph {et~al.}(2019)\citenamefont
  {Crawford}, \citenamefont {van Straaten}, \citenamefont {Wang}, \citenamefont
  {Parks}, \citenamefont {Campbell},\ and\ \citenamefont
  {Brierley}}]{Crawford2019}%
  \BibitemOpen
  \bibfield  {author} {\bibinfo {author} {\bibfnamefont {Ophelia}\ \bibnamefont
  {Crawford}}, \bibinfo {author} {\bibfnamefont {Barnaby}\ \bibnamefont {van
  Straaten}}, \bibinfo {author} {\bibfnamefont {Daochen}\ \bibnamefont {Wang}},
  \bibinfo {author} {\bibfnamefont {Thomas}\ \bibnamefont {Parks}}, \bibinfo
  {author} {\bibfnamefont {Earl}\ \bibnamefont {Campbell}}, \ and\ \bibinfo
  {author} {\bibfnamefont {Stephen}\ \bibnamefont {Brierley}},\ }\bibfield
  {title} {\enquote {\bibinfo {title} {Efficient quantum measurement of pauli
  operators},}\ }\href {https://arxiv.org/abs/1908.06942} {\bibfield  {journal}
  {\bibinfo  {journal} {arXiv:1908.06942}\ } (\bibinfo {year}
  {2019})}\BibitemShut {NoStop}%
\bibitem [{\citenamefont {Gokhale}\ and\ \citenamefont
  {Chong}(2019)}]{Gokhale2019-2}%
  \BibitemOpen
  \bibfield  {author} {\bibinfo {author} {\bibfnamefont {Pranav}\ \bibnamefont
  {Gokhale}}\ and\ \bibinfo {author} {\bibfnamefont {Frederic~T}\ \bibnamefont
  {Chong}},\ }\bibfield  {title} {\enquote {\bibinfo {title} {$o(n^3)$
  measurement cost for variational quantum eigensolver on molecular
  hamiltonians},}\ }\href {https://arxiv.org/abs/1908.11857} {\bibfield
  {journal} {\bibinfo  {journal} {arXiv:1908.11857}\ } (\bibinfo {year}
  {2019})}\BibitemShut {NoStop}%
\bibitem [{\citenamefont {Huggins}\ \emph {et~al.}(2019)\citenamefont
  {Huggins}, \citenamefont {McClean}, \citenamefont {Rubin}, \citenamefont
  {Jiang}, \citenamefont {Wiebe}, \citenamefont {Whaley},\ and\ \citenamefont
  {Babbush}}]{huggins2019efficient}%
  \BibitemOpen
  \bibfield  {author} {\bibinfo {author} {\bibfnamefont {William~J}\
  \bibnamefont {Huggins}}, \bibinfo {author} {\bibfnamefont {Jarrod}\
  \bibnamefont {McClean}}, \bibinfo {author} {\bibfnamefont {Nicholas}\
  \bibnamefont {Rubin}}, \bibinfo {author} {\bibfnamefont {Zhang}\ \bibnamefont
  {Jiang}}, \bibinfo {author} {\bibfnamefont {Nathan}\ \bibnamefont {Wiebe}},
  \bibinfo {author} {\bibfnamefont {K~Birgitta}\ \bibnamefont {Whaley}}, \ and\
  \bibinfo {author} {\bibfnamefont {Ryan}\ \bibnamefont {Babbush}},\ }\bibfield
   {title} {\enquote {\bibinfo {title} {Efficient and noise resilient
  measurements for quantum chemistry on near-term quantum computers},}\ }\href
  {https://arxiv.org/abs/1907.13117} {\bibfield  {journal} {\bibinfo  {journal}
  {arXiv:1907.13117}\ } (\bibinfo {year} {2019})}\BibitemShut {NoStop}%
\bibitem [{\citenamefont {Wecker}\ \emph {et~al.}(2015)\citenamefont {Wecker},
  \citenamefont {Hastings},\ and\ \citenamefont {Troyer}}]{troyer2015}%
  \BibitemOpen
  \bibfield  {author} {\bibinfo {author} {\bibfnamefont {Dave}\ \bibnamefont
  {Wecker}}, \bibinfo {author} {\bibfnamefont {Matthew~B}\ \bibnamefont
  {Hastings}}, \ and\ \bibinfo {author} {\bibfnamefont {Matthias}\ \bibnamefont
  {Troyer}},\ }\bibfield  {title} {\enquote {\bibinfo {title} {Progress towards
  practical quantum variational algorithms},}\ }\href {\doibase
  10.1103/PhysRevA.92.042303} {\bibfield  {journal} {\bibinfo  {journal} {Phys.
  Rev. A}\ }\textbf {\bibinfo {volume} {92}},\ \bibinfo {pages} {042303}
  (\bibinfo {year} {2015})}\BibitemShut {NoStop}%
\bibitem [{\citenamefont {Verdon}\ \emph
  {et~al.}(2019{\natexlab{b}})\citenamefont {Verdon}, \citenamefont
  {Broughton}, \citenamefont {McClean}, \citenamefont {Sung}, \citenamefont
  {Babbush}, \citenamefont {Jiang}, \citenamefont {Neven},\ and\ \citenamefont
  {Mohseni}}]{verdon2019learning}%
  \BibitemOpen
  \bibfield  {author} {\bibinfo {author} {\bibfnamefont {Guillaume}\
  \bibnamefont {Verdon}}, \bibinfo {author} {\bibfnamefont {Michael}\
  \bibnamefont {Broughton}}, \bibinfo {author} {\bibfnamefont {Jarrod~R}\
  \bibnamefont {McClean}}, \bibinfo {author} {\bibfnamefont {Kevin~J}\
  \bibnamefont {Sung}}, \bibinfo {author} {\bibfnamefont {Ryan}\ \bibnamefont
  {Babbush}}, \bibinfo {author} {\bibfnamefont {Zhang}\ \bibnamefont {Jiang}},
  \bibinfo {author} {\bibfnamefont {Hartmut}\ \bibnamefont {Neven}}, \ and\
  \bibinfo {author} {\bibfnamefont {Masoud}\ \bibnamefont {Mohseni}},\
  }\bibfield  {title} {\enquote {\bibinfo {title} {Learning to learn with
  quantum neural networks via classical neural networks},}\ }\href
  {https://arxiv.org/abs/1907.05415} {\bibfield  {journal} {\bibinfo  {journal}
  {arXiv:1907.05415}\ } (\bibinfo {year} {2019}{\natexlab{b}})}\BibitemShut
  {NoStop}%
\bibitem [{\citenamefont {Wilson}\ \emph {et~al.}(2019)\citenamefont {Wilson},
  \citenamefont {Stromswold}, \citenamefont {Wudarski}, \citenamefont
  {Hadfield}, \citenamefont {Tubman},\ and\ \citenamefont
  {Rieffel}}]{wilson2019optimizing}%
  \BibitemOpen
  \bibfield  {author} {\bibinfo {author} {\bibfnamefont {Max}\ \bibnamefont
  {Wilson}}, \bibinfo {author} {\bibfnamefont {Sam}\ \bibnamefont
  {Stromswold}}, \bibinfo {author} {\bibfnamefont {Filip}\ \bibnamefont
  {Wudarski}}, \bibinfo {author} {\bibfnamefont {Stuart}\ \bibnamefont
  {Hadfield}}, \bibinfo {author} {\bibfnamefont {Norm~M}\ \bibnamefont
  {Tubman}}, \ and\ \bibinfo {author} {\bibfnamefont {Eleanor}\ \bibnamefont
  {Rieffel}},\ }\bibfield  {title} {\enquote {\bibinfo {title} {Optimizing
  quantum heuristics with meta-learning},}\ }\href
  {https://arxiv.org/abs/1908.03185} {\bibfield  {journal} {\bibinfo  {journal}
  {arXiv:1908.03185}\ } (\bibinfo {year} {2019})}\BibitemShut {NoStop}%
\bibitem [{\citenamefont {Nakanishi}\ \emph {et~al.}(2019)\citenamefont
  {Nakanishi}, \citenamefont {Fujii},\ and\ \citenamefont
  {Todo}}]{nakanishi2019}%
  \BibitemOpen
  \bibfield  {author} {\bibinfo {author} {\bibfnamefont {Ken~M}\ \bibnamefont
  {Nakanishi}}, \bibinfo {author} {\bibfnamefont {Keisuke}\ \bibnamefont
  {Fujii}}, \ and\ \bibinfo {author} {\bibfnamefont {Synge}\ \bibnamefont
  {Todo}},\ }\bibfield  {title} {\enquote {\bibinfo {title} {Sequential minimal
  optimization for quantum-classical hybrid algorithms},}\ }\href
  {https://arxiv.org/abs/1903.12166} {\bibfield  {journal} {\bibinfo  {journal}
  {arXiv:1903.12166}\ } (\bibinfo {year} {2019})}\BibitemShut {NoStop}%
\bibitem [{\citenamefont {Parrish}\ \emph {et~al.}(2019)\citenamefont
  {Parrish}, \citenamefont {Iosue}, \citenamefont {Ozaeta},\ and\ \citenamefont
  {McMahon}}]{parrish2019}%
  \BibitemOpen
  \bibfield  {author} {\bibinfo {author} {\bibfnamefont {Robert~M}\
  \bibnamefont {Parrish}}, \bibinfo {author} {\bibfnamefont {Joseph~T}\
  \bibnamefont {Iosue}}, \bibinfo {author} {\bibfnamefont {Asier}\ \bibnamefont
  {Ozaeta}}, \ and\ \bibinfo {author} {\bibfnamefont {Peter~L}\ \bibnamefont
  {McMahon}},\ }\bibfield  {title} {\enquote {\bibinfo {title} {A {J}acobi
  diagonalization and {A}nderson acceleration algorithm for variational quantum
  algorithm parameter optimization},}\ }\href
  {https://arxiv.org/abs/1904.03206} {\bibfield  {journal} {\bibinfo  {journal}
  {arXiv:1904.03206}\ } (\bibinfo {year} {2019})}\BibitemShut {NoStop}%
\bibitem [{\citenamefont {Stokes}\ \emph {et~al.}(2019)\citenamefont {Stokes},
  \citenamefont {Izaac}, \citenamefont {Killoran},\ and\ \citenamefont
  {Carleo}}]{stokes2019quantum}%
  \BibitemOpen
  \bibfield  {author} {\bibinfo {author} {\bibfnamefont {James}\ \bibnamefont
  {Stokes}}, \bibinfo {author} {\bibfnamefont {Josh}\ \bibnamefont {Izaac}},
  \bibinfo {author} {\bibfnamefont {Nathan}\ \bibnamefont {Killoran}}, \ and\
  \bibinfo {author} {\bibfnamefont {Giuseppe}\ \bibnamefont {Carleo}},\
  }\bibfield  {title} {\enquote {\bibinfo {title} {Quantum natural gradient},}\
  }\href {https://arxiv.org/abs/1909.02108} {\bibfield  {journal} {\bibinfo
  {journal} {arXiv:1909.02108}\ } (\bibinfo {year} {2019})}\BibitemShut
  {NoStop}%
\bibitem [{\citenamefont {K{\"u}bler}\ \emph {et~al.}(2019)\citenamefont
  {K{\"u}bler}, \citenamefont {Arrasmith}, \citenamefont {Cincio},\ and\
  \citenamefont {Coles}}]{kubler2019adaptive}%
  \BibitemOpen
  \bibfield  {author} {\bibinfo {author} {\bibfnamefont {Jonas~M}\ \bibnamefont
  {K{\"u}bler}}, \bibinfo {author} {\bibfnamefont {Andrew}\ \bibnamefont
  {Arrasmith}}, \bibinfo {author} {\bibfnamefont {Lukasz}\ \bibnamefont
  {Cincio}}, \ and\ \bibinfo {author} {\bibfnamefont {Patrick~J}\ \bibnamefont
  {Coles}},\ }\bibfield  {title} {\enquote {\bibinfo {title} {An adaptive
  optimizer for measurement-frugal variational algorithms},}\ }\href
  {https://arxiv.org/abs/1909.09083} {\bibfield  {journal} {\bibinfo  {journal}
  {arXiv preprint arXiv:1909.09083}\ } (\bibinfo {year} {2019})}\BibitemShut
  {NoStop}%
\bibitem [{\citenamefont {Sweke}\ \emph {et~al.}(2019)\citenamefont {Sweke},
  \citenamefont {Wilde}, \citenamefont {Meyer}, \citenamefont {Schuld},
  \citenamefont {F{\"a}hrmann}, \citenamefont {Meynard-Piganeau},\ and\
  \citenamefont {Eisert}}]{sweke2019}%
  \BibitemOpen
  \bibfield  {author} {\bibinfo {author} {\bibfnamefont {Ryan}\ \bibnamefont
  {Sweke}}, \bibinfo {author} {\bibfnamefont {Frederik}\ \bibnamefont {Wilde}},
  \bibinfo {author} {\bibfnamefont {Johannes}\ \bibnamefont {Meyer}}, \bibinfo
  {author} {\bibfnamefont {Maria}\ \bibnamefont {Schuld}}, \bibinfo {author}
  {\bibfnamefont {Paul~K}\ \bibnamefont {F{\"a}hrmann}}, \bibinfo {author}
  {\bibfnamefont {Barth{\'e}l{\'e}my}\ \bibnamefont {Meynard-Piganeau}}, \ and\
  \bibinfo {author} {\bibfnamefont {Jens}\ \bibnamefont {Eisert}},\ }\bibfield
  {title} {\enquote {\bibinfo {title} {Stochastic gradient descent for hybrid
  quantum-classical optimization},}\ }\href {https://arxiv.org/abs/1910.01155}
  {\bibfield  {journal} {\bibinfo  {journal} {arXiv preprint arXiv:1910.01155}\
  } (\bibinfo {year} {2019})}\BibitemShut {NoStop}%
\bibitem [{\citenamefont {Zhang}\ and\ \citenamefont
  {Yin}(2019)}]{zhang2019collective}%
  \BibitemOpen
  \bibfield  {author} {\bibinfo {author} {\bibfnamefont {Dan-Bo}\ \bibnamefont
  {Zhang}}\ and\ \bibinfo {author} {\bibfnamefont {Tao}\ \bibnamefont {Yin}},\
  }\bibfield  {title} {\enquote {\bibinfo {title} {Collective optimization for
  variational quantum eigensolvers},}\ }\href
  {https://arxiv.org/abs/1910.14030} {\bibfield  {journal} {\bibinfo  {journal}
  {arXiv preprint arXiv:1910.14030}\ } (\bibinfo {year} {2019})}\BibitemShut
  {NoStop}%
\bibitem [{\citenamefont {Koczor}\ and\ \citenamefont
  {Benjamin}(2019)}]{koczor2019quantum}%
  \BibitemOpen
  \bibfield  {author} {\bibinfo {author} {\bibfnamefont {B{\'a}lint}\
  \bibnamefont {Koczor}}\ and\ \bibinfo {author} {\bibfnamefont {Simon~C}\
  \bibnamefont {Benjamin}},\ }\bibfield  {title} {\enquote {\bibinfo {title}
  {Quantum natural gradient generalised to non-unitary circuits},}\ }\href
  {https://arxiv.org/abs/1912.08660} {\bibfield  {journal} {\bibinfo  {journal}
  {arXiv preprint arXiv:1912.08660}\ } (\bibinfo {year} {2019})}\BibitemShut
  {NoStop}%
\bibitem [{\citenamefont {Lavrijsen}\ \emph {et~al.}(2020)\citenamefont
  {Lavrijsen}, \citenamefont {Tudor}, \citenamefont {M{\"u}ller}, \citenamefont
  {Iancu},\ and\ \citenamefont {de~Jong}}]{lavrijsen2020classical}%
  \BibitemOpen
  \bibfield  {author} {\bibinfo {author} {\bibfnamefont {Wim}\ \bibnamefont
  {Lavrijsen}}, \bibinfo {author} {\bibfnamefont {Ana}\ \bibnamefont {Tudor}},
  \bibinfo {author} {\bibfnamefont {Juliane}\ \bibnamefont {M{\"u}ller}},
  \bibinfo {author} {\bibfnamefont {Costin}\ \bibnamefont {Iancu}}, \ and\
  \bibinfo {author} {\bibfnamefont {Wibe}\ \bibnamefont {de~Jong}},\ }\bibfield
   {title} {\enquote {\bibinfo {title} {Classical optimizers for noisy
  intermediate-scale quantum devices},}\ }\href
  {https://arxiv.org/abs/2004.03004} {\bibfield  {journal} {\bibinfo  {journal}
  {arXiv preprint arXiv:2004.03004}\ } (\bibinfo {year} {2020})}\BibitemShut
  {NoStop}%
\bibitem [{\citenamefont {Harrow}\ and\ \citenamefont
  {Napp}(2019)}]{harrow2019}%
  \BibitemOpen
  \bibfield  {author} {\bibinfo {author} {\bibfnamefont {Aram}\ \bibnamefont
  {Harrow}}\ and\ \bibinfo {author} {\bibfnamefont {John}\ \bibnamefont
  {Napp}},\ }\bibfield  {title} {\enquote {\bibinfo {title} {Low-depth gradient
  measurements can improve convergence in variational hybrid quantum-classical
  algorithms},}\ }\href {https://arxiv.org/abs/1901.05374} {\bibfield
  {journal} {\bibinfo  {journal} {arXiv:1901.05374}\ } (\bibinfo {year}
  {2019})}\BibitemShut {NoStop}%
\bibitem [{\citenamefont {Kingma}\ and\ \citenamefont {Ba}(2015)}]{Kingma2015}%
  \BibitemOpen
  \bibfield  {author} {\bibinfo {author} {\bibfnamefont {Diederik~P}\
  \bibnamefont {Kingma}}\ and\ \bibinfo {author} {\bibfnamefont {Jimmy}\
  \bibnamefont {Ba}},\ }\bibfield  {title} {\enquote {\bibinfo {title} {Adam:
  {A} method for stochastic optimization},}\ }in\ \href
  {http://arxiv.org/abs/1412.6980} {\emph {\bibinfo {booktitle} {Proceedings of
  the 3rd International Conference on Learning Representations (ICLR)}}}\
  (\bibinfo {year} {2015})\BibitemShut {NoStop}%
\bibitem [{\citenamefont {Campbell}(2019)}]{campbell2019random}%
  \BibitemOpen
  \bibfield  {author} {\bibinfo {author} {\bibfnamefont {Earl}\ \bibnamefont
  {Campbell}},\ }\bibfield  {title} {\enquote {\bibinfo {title} {Random
  compiler for fast hamiltonian simulation},}\ }\href@noop {} {\bibfield
  {journal} {\bibinfo  {journal} {Physical review letters}\ }\textbf {\bibinfo
  {volume} {123}},\ \bibinfo {pages} {070503} (\bibinfo {year}
  {2019})}\BibitemShut {NoStop}%
\bibitem [{\citenamefont {Rubin}\ \emph {et~al.}(2018)\citenamefont {Rubin},
  \citenamefont {Babbush},\ and\ \citenamefont
  {McClean}}]{rubin2018application}%
  \BibitemOpen
  \bibfield  {author} {\bibinfo {author} {\bibfnamefont {Nicholas~C}\
  \bibnamefont {Rubin}}, \bibinfo {author} {\bibfnamefont {Ryan}\ \bibnamefont
  {Babbush}}, \ and\ \bibinfo {author} {\bibfnamefont {Jarrod}\ \bibnamefont
  {McClean}},\ }\bibfield  {title} {\enquote {\bibinfo {title} {Application of
  fermionic marginal constraints to hybrid quantum algorithms},}\ }\href
  {https://iopscience.iop.org/article/10.1088/1367-2630/aab919/meta} {\bibfield
   {journal} {\bibinfo  {journal} {New Journal of Physics}\ }\textbf {\bibinfo
  {volume} {20}},\ \bibinfo {pages} {053020} (\bibinfo {year}
  {2018})}\BibitemShut {NoStop}%
\bibitem [{\citenamefont {Kandala}\ \emph {et~al.}(2017)\citenamefont
  {Kandala}, \citenamefont {Mezzacapo}, \citenamefont {Temme}, \citenamefont
  {Takita}, \citenamefont {Brink}, \citenamefont {Chow},\ and\ \citenamefont
  {Gambetta}}]{kandala2017}%
  \BibitemOpen
  \bibfield  {author} {\bibinfo {author} {\bibfnamefont {Abhinav}\ \bibnamefont
  {Kandala}}, \bibinfo {author} {\bibfnamefont {Antonio}\ \bibnamefont
  {Mezzacapo}}, \bibinfo {author} {\bibfnamefont {Kristan}\ \bibnamefont
  {Temme}}, \bibinfo {author} {\bibfnamefont {Maika}\ \bibnamefont {Takita}},
  \bibinfo {author} {\bibfnamefont {Markus}\ \bibnamefont {Brink}}, \bibinfo
  {author} {\bibfnamefont {Jerry~M}\ \bibnamefont {Chow}}, \ and\ \bibinfo
  {author} {\bibfnamefont {Jay~M}\ \bibnamefont {Gambetta}},\ }\bibfield
  {title} {\enquote {\bibinfo {title} {Hardware-efficient variational quantum
  eigensolver for small molecules and quantum magnets},}\ }\href {\doibase
  10.1038/nature23879} {\bibfield  {journal} {\bibinfo  {journal} {Nature}\
  }\textbf {\bibinfo {volume} {549}},\ \bibinfo {pages} {242} (\bibinfo {year}
  {2017})}\BibitemShut {NoStop}%
\bibitem [{IBM(2018)}]{IBMQ14}%
  \BibitemOpen
  \href@noop {} {\enquote {\bibinfo {title} {{IBM} {Q} 16 {M}elbourne backend
  specification},}\ }\bibinfo {howpublished}
  {\url{https://github.com/Qiskit/ibmq-device-information/tree/master/backends/melbourne/V1}}
  (\bibinfo {year} {2018})\BibitemShut {NoStop}%
\bibitem [{\citenamefont {et.al.}(2019)}]{gadi_aleksandrowicz_2019_2562111}%
  \BibitemOpen
  \bibfield  {author} {\bibinfo {author} {\bibfnamefont {Gadi~Aleksandrowicz}\
  \bibnamefont {et.al.}},\ }\href {\doibase 10.5281/zenodo.2562111} {\enquote
  {\bibinfo {title} {{Qiskit: An Open-source Framework for Quantum
  Computing}},}\ } (\bibinfo {year} {2019})\BibitemShut {NoStop}%
\end{thebibliography}
\end{document}